\documentclass[english,11pt]{article}
\usepackage[T1]{fontenc}
\usepackage[latin9]{inputenc}
\usepackage{geometry}
\usepackage{amsmath}
\usepackage{amsthm}
\usepackage{amssymb}
\usepackage{graphicx}
\usepackage{float}
\usepackage{eurosym}
\usepackage{xspace}
\usepackage[shortlabels]{enumitem}

  \theoremstyle{remark}
  \newtheorem{rem}{\protect\remarkname}
  \theoremstyle{plain}
  
  \theoremstyle{plain}
  
  \theoremstyle{plain}
  \newtheorem*{theo*}{\protect\theoremname}
  \theoremstyle{definition}
  
  \theoremstyle{plain}
  
\theoremstyle{plain}

\usepackage{babel}
  \providecommand{\definitionname}{Definition}
  \providecommand{\propositionname}{Proposition}
  \providecommand{\theoremname}{Theorem}
  \providecommand{\remarkname}{Remark}
\providecommand{\corollaryname}{Corollary}
\providecommand{\lemmaname}{Lemma}

\begin{document}

\title{Mid-price estimation for European corporate bonds: a particle filtering approach\thanks{This research has been conducted with the support of the Research Initiative ``Mod\'elisation des march\'es actions, obligations et d\'eriv\'es''  financed by HSBC France under the aegis of the Europlace Institute of Finance. However, the ideas presented in this paper do not reflect the views or practice at HSBC. The authors would like to thank Paul-Henry Bacher~(HSBC), Nicolas Chopin (ENSAE-CREST), Nicolas Grandchamp des Raux~(HSBC), Alexandre Guignot~(HSBC), Jean-Michel Lasry (Institut Louis Bachelier), Guillaume Macey (HSBC), and S\'ebastien Roland~(HSBC) for the conversations they had on the subject. Binbin Zhang (Institut Europlace de Finance) also deserves to be warmly thanked for her help and her insightful ideas. Finally, an anonymous referee deserves to be warmly thanked for his insightful remarks and his thorough reading of our paper.}}
\date{}
\author{Olivier Gu\'eant\thanks{Universit\'e Paris 1 Panth\'eon-Sorbonne. Centre d'Economie de la Sorbonne. 106, Bd de l'H\^opital, 75013 Paris.}, Jiang Pu\thanks{Institut Europlace de Finance. 28, Place de la Bourse, 75002 Paris.}}
\maketitle
\begin{abstract}
In most illiquid markets, there is no obvious proxy for the market price of an asset. The European corporate bond market is an archetypal example of such an illiquid market where mid-prices can only be estimated with a statistical model. In this OTC market, dealers / market makers only have access, indeed, to partial information about the market. In real time, they know the price associated with their trades on the dealer-to-dealer (D2D) and dealer-to-client (D2C) markets, they know the result of the requests for quotes (RFQ) they answered, and they have access to composite prices (\emph{e.g.}, Bloomberg CBBT). This paper presents a Bayesian method for estimating the mid-price of corporate bonds by using the real-time information available to a dealer. This method relies on recent ideas coming from the particle filtering~/~sequential Monte-Carlo literature.

\vspace{5mm}

\vspace{5mm}
\end{abstract}

\section{Introduction}

The European market for corporate bonds is experiencing important changes. Because of a long-lasting low-interest rate environment, the amount of bond issuance have largely increased and the overall size of the market has subsequently skyrocketed (see \cite{icma} and \cite{iosco}). Furthermore, although it remains largely OTC, the European market for corporate bonds is relying more and more on electronic platforms (mainly multi-dealer-to-client platforms -- MD2C). As a consequence, the amount of data available to sell-siders (\emph{i.e.} dealers / market makers) has been increasing and that has opened the door to statistical modeling, especially for building or improving market making algorithms.\\

The role of market makers in any market is to provide liquidity to other market participants, \emph{i.e.} proposing prices at which they stand ready to buy and sell a wide variety of assets. In the case of corporate bond markets, market makers have traditionally played a major role in the liquidity provision and price formation processes. Nowadays, there is however an important debate about the role played by market makers in corporate bond markets. On the one hand, market makers want to hold less risk on their balance sheet. Subsequently, sales departments of most investment banks have invested in data analysis to better anticipate clients' needs and to be proactive in order to accelerate the turnover of bonds, and there is even, especially in the US, a move from the traditional principal trading model towards a riskless principal trading one in which market makers try to directly match interests (see for instance \cite{harris}). On the other hand, market makers are investing time and resources in automating the market making process through algorithms.\\

The problem faced by market makers is in fact a complex one from a quantitative viewpoint with both static and dynamic components. First, any market maker faces a static optimization problem regarding the width of the spread he quotes: tight spread and numerous trades \emph{versus} large spread and few trades. Second, and foremost, any market maker has to solve a dynamic optimization problem to adapt his bid and ask quotes to his inventory as the main risk he faces is that of holding assets while their price is moving, possibly adversely. This complex problem has been studied by both economists (see for instance \cite{grossman1988liquidity} or the old papers \cite{ho1981optimal, ho1983dynamics}) and mathematicians (see for instance \cite{avellaneda2008high}, \cite{cartea2013risk}, \cite{cartea2015algorithmic}, \cite{gueant2013dealing}, \cite{gueant2017optimal}, \cite{gueantbook}, etc.). In all the quantitative models developed to solve the market making problem, optimal bid and offer quotes are derived with respect to a benchmark price, which is, depending on the wording of the authors of the model, a fair price, a mid-price, or a reference price. In all cases, the associated price process is exogenous. The role of a quantitative market making model is indeed not to give a price to securities, but instead to decide on the margin / markup a market maker should set in order to make money while mitigating the inventory risk, \emph{i.e.}, the risk of being unable to unwind (at low cost) a position in adverse market conditions.\\

In many markets, there are natural benchmark prices. In the case of liquid markets with limit order books, the mid-price or a volume-weighted average of the best bid and ask prices are indeed natural and relevant. However, in the case of illiquid OTC markets, the situation is not the same. Corporate bond markets are archetypal of such markets where there is no natural reference price. Most practitioners consider the CBBT mid-price provided by Bloomberg as a reference price, but it is clearly a choice by default, especially since the algorithm used to obtain the value of the CBBT bid and offer prices is not public and sometimes fails to provide prices, for instance because the calculated bid price turns out to be above the calculated ask price.\footnote{See Bloomberg documentation on CBBT.}\\

In this paper, we aim at building a reference price by using the relevant information available to market makers. Our focus is on the European market for corporate bonds where market makers know in real time the price associated with their trades on the dealer-to-dealer (D2D) and dealer-to-client (D2C) markets and the result of the requests for quotes (RFQ) they answered. Of course, they also observe in real time the CBBT, but our goal is to build an alternative estimator.\footnote{In the model we present, the CBBT is only used for estimating parameters.} \footnote{It is noteworthy that, at the time of finishing this paper (2018), post-trade transparency just started to be enforced in the European Union (through MiFID II). We believe that there will be one or two years before post-trade transparency data is widely used in a relevant way for business purposes. An interesting parallel can be drawn with the use of TRACE data in the US (see \cite{trace}). As far as post-trade transparency data is concerned, it must be noted that there is always a lag between the time of the trade and the time at which it is made public. However, our algorithm could be generalized to account for this kind of data.}\\

The academic literature on mid-price estimation in the European corporate bond market is non-existent, mainly because of the lack of data. Practitioners used to consider the CBBT as a proxy for the mid-price, because there was almost no other choice. Today, however, with the increasing importance of electronic trading, and especially with the increasing number of RFQs received, answered, and recorded by dealers,\footnote{RFQ data from a major investment bank has been used in \cite{fermanian} for analyzing the behavior of dealers and clients on MD2C platforms through a statistical model.} alternatives to CBBT prices can be considered. RFQs contain indeed information about the transactions missed by a requested dealer when he did not propose the best price to the client. For using this type of nonlinear (censored) information, Kalman-like filters are not enough. Consequently, we propose a method based on Bayesian filtering, and more exactly on a particle filtering (PF) / sequential Monte-Carlo (SMC) approach. It provides a distribution for the reference yield to benchmark of each considered bond\footnote{Throughout this article, we work in terms of yield to benchmark (difference between the yield to maturity of a bond and that of its natural benchmark -- typically a government bond with similar duration) rather than price in order to eliminate the risk-free rate component in bond prices.} -- a reference yield to benchmark being defined somehow endogenously in the model by the property that ``the probability to observe buy or sell trades at any given distance from the reference yield to benchmark only depends on that distance and not on the side'' (at least when market makers' inventories are flat on average).\\

Particle filtering / sequential Monte-Carlo is a class of approaches and algorithms providing Monte Carlo
approximations of a sequence of probability distributions in a Bayesian framework. As in classical filtering, one wants to infer the probability distribution of a phenomenon that is not observed directly, but on which one receives partial data (due to censorship or noise) in a sequential manner. The main idea underlying these techniques is to model the phenomenon by a Markov chain and to approximate distributions through a cloud of weighted random samples that is propagated over time using the dynamics of the Markov chain and the application of Bayes' rule to the observations.\\

These approaches are often only restricted to low-dimensional problems. An example of application of this type of approaches is described in our paper \cite{gpindices} on credit indices in which we use the doubly-Bayesian  $\textrm{SMC}^2$ approach of Chopin \emph{et al.} \cite{smc2}. In the high-dimensional case of corporate bonds, the ``curse of dimensionality'' that traditionally affects particle filtering / sequential Monte-Carlo approaches has to be put in perspective. In illiquid markets, indeed, because transactions seldom occur simultaneously on several securities and because prices diffuse, this type of approaches seems to scale far better than expected.\\

In Section 2, we first present the principle of our particle filtering algorithm. We also derive the recursive equations and then detail the simulation process associated with them. In Section 3, we discuss the estimation of the parameters and we illustrate our algorithm on a few European corporate bonds. In Section~4, we discuss the advantages and drawbacks of our approach and the numerous possible generalizations.\\

\section{Principle of the algorithm}

\subsection{Modeling framework and goals}

Let $\left(\Omega,\mathcal{F},\left(\mathcal{F}_{t}\right)_{t\in\mathbb{R}_{+}},\mathbb{P}\right)$
be a filtered probability space, with $\left(\mathcal{F}_{t}\right)_{t\in\mathbb{R}_{+}}$
satisfying the usual conditions.\\

We consider a set of $d$ corporate bonds. Instead of considering bond prices, a common market practice (at least for investment grade bonds) consists in considering yields to benchmark (YtB) in order to remove the risk-free interest rate component of bond prices. In what follows, we model the mid-YtB of the $d$ corporate bonds by a $d$-dimensional process $(y_t)_t$ and we assume that
\begin{equation}\label{dS}
\forall i \in \{1, \ldots, d\}, dy^i_t = \sigma^i dW_t^i,
\end{equation}
where $\left(W_{t}\right)_t$ is a $d$-dimensional Brownian motion adapted to $\left(\mathcal{F}_{t}\right)_{t\in\mathbb{R}_{+}}$, with
$$d\left\langle W_{t}^{i},W_{t}^{j}\right\rangle =\rho^{i,j}dt.$$
We denote by $\Sigma$ the covariance matrix
\begin{eqnarray*}
\Sigma^{i,j} & = & \rho^{i,j}\sigma^{i}\sigma^{j}.
\end{eqnarray*}

\begin{rem}
The model we consider for the dynamics of the (mid-)YtB is of course simplistic. Our choice is driven by several considerations. First, we chose a simple model for the sake of presentation. Considering an Ornstein-Uhlenbeck is another possibility and the principle of the algorithm would be the same. Second, given the nature of the algorithm we consider, what matters is the diffusive character of the model (and the value of the ``volatility'' parameters $(\sigma^i)_i$) rather than the specific dynamics of the process.
\end{rem}

We introduce another process $(x_t)_t$ following a $d$-dimensional Ornstein-Ulhenbeck process:
\begin{equation}\label{dx}dx_t = -A x_t dt + V dB_t, \qquad x_0 \textrm{\; given},
\end{equation}
where $A$ and $V$ are $d\times d$ matrices, and $\left(B_{t}\right)_t$ a $d$-dimensional standard Brownian motion adapted to $\left(\mathcal{F}_{t}\right)_{t\in\mathbb{R}_{+}}$, assumed to be independent from the process $(W_t)_t$.\footnote{We could make the model more general by assuming a nontrivial correlation structure between $W$ and $B$, but the correlation structure would be very difficult to estimate.}\\

We recall (see for instance \cite{meucci}) that $(x_t)_t$ is a Gaussian process with, $\forall t \in \mathbb{R}_+, \forall \tau > 0$,
$$\mathbb{E}[x_{t+\tau}|x_t] = e^{-A\tau} x_t$$
and
$$\mathbb{V}[x_{t+\tau}|x_t] = \Gamma(\tau),$$
with $$\textrm{vec}(\Gamma(\tau)) = (A \otimes I_d + I_d \otimes A)^{-1}\left(I_{d^2} - \exp(-(A \otimes I_d + I_d \otimes A)\tau)\right) \textrm{vec}(VV'),$$ where $\textrm{vec}(\cdot)$ refers to the vectorization operator, \emph{i.e.}, $$\textrm{vec}((M^{i,j})_{1 \le i,j \le d}) = (M^{1,1}, \ldots, M^{d,1}, \ldots, M^{1,d}, \ldots, M^{d,d})'.$$

\begin{rem}
In this paper, we consider that $A$ is a diagonal matrix. This is important in terms of computation since $(A \otimes I_d + I_d \otimes A)^{-1}\left(I_{d^2} - \exp(-(A \otimes I_d + I_d \otimes A)\tau)\right)$ is then a diagonal matrix whereas in general it is a square matrix of dimension $d^2\times d^2$ (which can hardly be used when $d$ is large). It is noteworthy, however, that $V$ can be of any form. In practice, if $A = \textrm{diag}(a^1, \ldots, a^d)$, then we have $$\Gamma^{i,j}(\tau) = \frac{1}{a^i+a^j} \left(1 - e^{-(a^i + a^j) \tau}\right) (VV')^{i,j}.$$
\end{rem}

\begin{rem}
Instead of considering continuous paths for $(x_t)_t$, one could also consider that the random variables $(x_t)_t$ are i.i.d. (and Gaussian). This can be considered as a limit case of ours where $A=a I_d$ for $a$ large, and $V$ rescaled accordingly.
\end{rem}

For $i \in \{1, \ldots, d\}$, we introduce the half bid-ask spread process $(\psi^i_t)_t$ of asset $i$ (in terms of yield to benchmark) defined by
$$\psi^i_t = \Psi^i \exp(x^i_t),\qquad \Psi^i \textrm{\; given}.$$
In other words, $y^i_t + \psi^i_t$ and $y^i_t - \psi^i_t$ are respectively the bid-YtB and the ask-YtB.\footnote{The bid-YtB has to be higher than the ask-YtB for the bid price to be lower than the ask price.}\\

We consider here that the matrices $\Sigma$, $A$, $V$, and the vector $(\Psi^i)_{1 \le i \le d}$ are given. We will discuss estimation in Section 4.\\

We take the viewpoint of a given dealer $D$ and we consider that the information available to him corresponds to $5$ different situations:
\begin{itemize}
  \item[$J_t = (i,1)$:] A client buys bond $i$ from dealer $D$ at time $t$ (by voice or through a platform). In that case, we assume that the YtB associated with the transaction is $Y^i_t = y^i_t - \psi^i_t + \epsilon^i_t$ and the observation of dealer $D$ is $O_t = Y^i_t$, where $\epsilon^i_t$ is a random variable $\mathcal{N}(0,{\sigma^i_{\epsilon}}^2)$ assumed to be independent of all other random variables.
  \item[$J_t = (i,2)$:] A client sells bond $i$ to dealer $D$ at time $t$ (by voice or through a platform). In that case, we assume that the YtB associated with the transaction is $Y^i_t = y^i_t + \psi^i_t + \epsilon^i_t$ and the observation of dealer $D$ is $O_t = Y^i_t$, where $\epsilon^i_t$ is a random variable $\mathcal{N}(0,{\sigma^i_{\epsilon}}^2)$ assumed to be independent of all other random variables.
  \item[$J_t = (i,3)$:] A client buys bond $i$ from another dealer at time $t$ through a RFQ. We assume that dealer $D$ has proposed a YtB $Z^i_t$ but was not chosen because a better price was proposed by the other dealer. In that case, we assume that the YtB associated with the transaction is $Y^i_t = y^i_t - \psi^i_t + \epsilon^i_t$, where $\epsilon^i_t$ is a random variable $\mathcal{N}(0,{\sigma^i_{\epsilon}}^2)$ assumed to be independent of all other random variables. The observation of dealer $D$ is $O_t = 1_{Y^i_t \ge  Z^i_t}$.
  \item[$J_t = (i,4)$:] A client sells bond $i$ to another dealer at time $t$ through a RFQ. We assume that dealer $D$ has proposed a YtB $Z^i_t$ but was not chosen because a better price was proposed by the other dealer. In that case, we assume that the YtB associated with the transaction is $Y^i_t = y^i_t + \psi^i_t + \epsilon^i_t$, where $\epsilon^i_t$ is a random variable $\mathcal{N}(0,{\sigma^i_{\epsilon}}^2)$ assumed to be independent of all other random variables.  The observation of dealer $D$ is $O_t =  1_{Y^i_t \le  Z^i_t}$.
  \item[$J_t = (i,5)$:] Another dealer transacted bond $i$ with dealer $D$ on the inter-dealer broker (IDB) market. In that case, we assume that the YtB associated with the transaction, denoted by $Y^i_t$, is in the range $[y^i_t - \alpha^i_t + \epsilon^i_t, y^i_t + \alpha^i_t + \epsilon^i_t]$, where $\epsilon^i_t$ is a random variable $\mathcal{N}(0,{\sigma^i_{\epsilon}}^2)$ assumed to be independent of all other random variables, and where $\alpha^i_t$ can for instance be chosen proportional to $\psi^i_t$ or proportional to a typical size for the bid-ask spread (in terms of yield to benchmark) of asset $i$.\footnote{Here it may be useful to consider the CBBT bid-ask spread.} The observation of dealer $D$ is of course $O_t = Y^i_t$.
\end{itemize}

The first 2 cases are referred to as D2C trades. The next 2 are also D2C trades, but from the point of view of dealer $D$ they are ``Traded Away'' RFQs. The last case is a D2D trade.\\

Our main goal is to develop, from the above information, a mathematical method and an algorithm for the online estimation of the mid-YtB of the securities along with their (half) bid-ask spread. Our focus is on the live problem faced by market makers (or market making algorithms) when they have to stream or answer bid and/or offer YtB/prices for an asset. A slightly different -- but intimately related -- problem is the \emph{ex-post} estimation of mid-YtB and half bid-ask spread trajectories given observations over a chosen time interval (for instance a day of trading). In the former case, the estimation only depends on the past and the present, while in the latter, it depends on all the observations.\\

The method we propose is based on a sequential Monte-Carlo algorithm. The principle of the algorithm is to draw particles and to use importance sampling based on Bayes' formula in order to derive empirically the distribution of the mid-YtB and half bid-ask spread processes conditionally on the observations.\\

At time $\tau_0=0$, we consider a prior density $\pi(\cdot)$ for $(y_0,x_0)$ and we consider a sample $(y^0_{0,k},x^0_{0,k})_{1 \le k \le K}$ of size $K$ drawn from $\pi$.\footnote{Throughout the article, $K$ always stands for the number of particles and $k$ for the index of a particle.} We also define half bid-ask spread initial points by $$\forall k \in \{1, \ldots, K\}, \psi^0_{0,k} = \left(
                      \begin{array}{c}
                        \Psi^1 \exp(x^{0,1}_{0,k}) \\
                        \vdots \\
                        \Psi^d \exp(x^{0,d}_{0,k}) \\
                      \end{array}
                    \right).$$

Let us consider that observations correspond to times $\tau_1 <  \ldots < \tau_N$.\\

The output of our algorithm is a family of families of partial trajectories indexed by the letter~$m$ $$\left((y^m_{n,k},x^m_{n,k},\psi^m_{n,k})_{0 \le n \le m, 1 \le k \le K}\right)_{0 \le m \le N}.$$ For each $m \in \{ 0, \ldots, N\}$,  $\left(\left(y^m_{n,k},\psi^m_{n,k}\right)_{0 \le n \le m}\right)_{1 \le k \le K}$ represents a sample of size $K$ of independent trajectories of the $d$-dimensional mid-YtB and half bid-ask spread processes, at times $\tau_1 <  \ldots < \tau_m$.

\begin{rem}
It is important to note that, throughout this paper, we assume that the very fact of observing data does not yield any information about the trajectories of the YtB/price and half bid-ask spread processes (beyond the value of the observations themselves). In other words, everything happens as if the observation dates and types were drawn randomly, independently of the stochastic processes $(y_t)_t$ and $(\psi_t)_t$.\\
\end{rem}

\subsection{The mathematics of the sequential Monte-Carlo algorithm}

\subsubsection{Likelihood recursion: general equations}

The SMC/PF algorithm is based on a recursive reasoning. In order to present the principle of the approach, let us consider $m \in \{ 0, \ldots, N-1\}$. For going from time $\tau_m$ to time $\tau_{m+1}$, let us first denote by $i$ the first component of $J_{\tau_{m+1}}$ and by $\tilde{y}_{\tau_{m+1}}$ the random variable
$$\tilde{y}_{\tau_{m+1}} = \left(y^0_{\tau_{m+1}}, \ldots, y^{i - 1}_{\tau_{m+1}}, y^{i}_{\tau_{m+1}}+\epsilon^{i}_{\tau_{m+1}}, y^{i + 1}_{\tau_{m+1}}, \ldots, y^d_{\tau_{m+1}}\right).$$
In terms of likelihood, we have
\begin{eqnarray*}
&&p\left((y_{\tau_n}, \psi_{\tau_n})_{0\le n \le m+1}, y^{i}_{\tau_{m+1}}+\epsilon^{i}_{\tau_{m+1}} | (J_{\tau_n}, O_{\tau_n})_{1\le n \le m+1}\right)\\
&=& p\left(y^{i}_{\tau_{m+1}} | (y_{\tau_n}, \psi_{\tau_n})_{0\le n \le m}, \tilde{y}_{\tau_{m+1}}, \psi_{\tau_{m+1}}, (J_{\tau_n}, O_{\tau_n})_{1\le n \le m+1}\right)\\
&&\times p\left((y_{\tau_n}, \psi_{\tau_n})_{0\le n \le m}, \tilde{y}_{\tau_{m+1}}, \psi_{\tau_{m+1}} | (J_{\tau_n}, O_{\tau_n})_{1\le n \le m+1}\right)\\
&=& p\left(y^{i}_{\tau_{m+1}} | y^{i}_{\tau_m}, y^{i}_{\tau_{m+1}}+\epsilon^{i}_{\tau_{m+1}}\right)\\
&&\times p\left((y_{\tau_n}, \psi_{\tau_n})_{0\le n \le m}, \tilde{y}_{\tau_{m+1}}, \psi_{\tau_{m+1}} | (J_{\tau_n}, O_{\tau_n})_{1\le n \le m+1}\right)\\
\end{eqnarray*}
Therefore, in order to draw in the distribution of
$$(y_{\tau_n}, \psi_{\tau_n})_{0\le n \le m+1} | (J_{\tau_n}, O_{\tau_n})_{1\le n \le m+1}$$
or in the extended distribution of
$$(y_{\tau_n}, \psi_{\tau_n})_{0\le n \le m+1}, y^{i}_{\tau_{m+1}}+\epsilon^{i}_{\tau_{m+1}} | (J_{\tau_n}, O_{\tau_n})_{1\le n \le m+1}$$
we shall first draw in the distribution of
$$(y_{\tau_n}, \psi_{\tau_n})_{0\le n \le m}, \tilde{y}_{\tau_{m+1}}, \psi_{\tau_{m+1}} | (J_{\tau_n}, O_{\tau_n})_{1\le n \le m+1}$$
and then draw $y^{i}_{\tau_{m+1}}$ conditionally on $y^{i}_{\tau_m}$ and $y^{i}_{\tau_{m+1}}+\epsilon^{i}_{\tau_{m+1}}$.
For the latter, we know from classical Gaussian filtering, or from a trivial application of Bayes' rule, that
\begin{equation}
\label{cond}y^{i}_{\tau_{m+1}} | y^{i}_{\tau_m}, y^{i}_{\tau_{m+1}}+\epsilon^{i}_{\tau_{m+1}} \sim \mathcal{N}\left(\frac{{\sigma^{i}}^2 (\tau_{m+1} - \tau_m) (y^{i}_{\tau_{m+1}}+\epsilon^{i}_{\tau_{m+1}}) + {\sigma^{i}_{\epsilon}}^2 y^{i}_{\tau_m} }{{\sigma^{i}}^2 (\tau_{m+1} - \tau_m) + {\sigma^{i}_{\epsilon}}^2}, \frac{{\sigma^{i}}^2 (\tau_{m+1} - \tau_m) {\sigma^{i}_{\epsilon}}^2 }{{\sigma^{i}}^2 (\tau_{m+1} - \tau_m) + {\sigma^{i}_{\epsilon}}^2}  \right)
\end{equation}
For drawing in the distribution
$$p\left((y_{\tau_n}, \psi_{\tau_n})_{0\le n \le m}, \tilde{y}_{\tau_{m+1}}, \psi_{\tau_{m+1}} | (J_{\tau_n}, O_{\tau_n})_{1\le n \le m+1}\right)$$
we use a recursive reasoning. For that purpose, we  use Bayes' rule:
\begin{eqnarray*}
  % \nonumber to remove numbering (before each equation)
    &&p\left((y_{\tau_n}, \psi_{\tau_n})_{0\le n \le m}, \tilde{y}_{\tau_{m+1}}, \psi_{\tau_{m+1}} | (J_{\tau_n}, O_{\tau_n})_{1\le n \le m+1}\right)\\
     &=& \frac{p\left( (J_{\tau_n}, O_{\tau_n})_{1\le n \le m+1} | (y_{\tau_n}, \psi_{\tau_n})_{0\le n \le m}, \tilde{y}_{\tau_{m+1}}, \psi_{\tau_{m+1}} \right) p\left((y_{\tau_n}, \psi_{\tau_n})_{0\le n \le m}, \tilde{y}_{\tau_{m+1}}, \psi_{\tau_{m+1}} \right)}{p\left((J_{\tau_n}, O_{\tau_n})_{1\le n \le m+1}\right)} \\
     &=&p\left( J_{\tau_{m+1}}, O_{\tau_{m+1}} | (J_{\tau_n}, O_{\tau_n})_{1\le n \le m}, (y_{\tau_n}, \psi_{\tau_n})_{0\le n \le m}, \tilde{y}_{\tau_{m+1}}, \psi_{\tau_{m+1}}\right)\\
     &&\times \frac{ p\left( (J_{\tau_n}, O_{\tau_n})_{1\le n \le m} | (y_{\tau_n}, \psi_{\tau_n})_{0\le n \le m}, \tilde{y}_{\tau_{m+1}}, \psi_{\tau_{m+1}}\right) p\left((y_{\tau_n}, \psi_{\tau_n})_{0\le n \le m},\tilde{y}_{\tau_{m+1}}, \psi_{\tau_{m+1}}\right)}{p\left(J_{\tau_{m+1}}, O_{\tau_{m+1}} | (J_{\tau_n}, O_{\tau_n})_{1\le n \le m} \right)p\left((J_{\tau_n}, O_{\tau_n})_{1\le n \le m}\right)}.\\
\end{eqnarray*}
We clearly have
$$p\left( (J_{\tau_n}, O_{\tau_n})_{1\le n \le m} | (y_{\tau_n}, \psi_{\tau_n})_{0\le n \le m}, \tilde{y}_{\tau_{m+1}}, \psi_{\tau_{m+1}}\right) = p\left( (J_{\tau_n}, O_{\tau_n})_{1\le n \le m} | (y_{\tau_n}, \psi_{\tau_n})_{0\le n \le m}\right).$$
Therefore,
\begin{eqnarray*}
  % \nonumber to remove numbering (before each equation)
    &&p\left((y_{\tau_n}, \psi_{\tau_n})_{0\le n \le m}, \tilde{y}_{\tau_{m+1}}, \psi_{\tau_{m+1}} | (J_{\tau_n}, O_{\tau_n})_{1\le n \le m+1}\right)\\
     &=&p\left( J_{\tau_{m+1}}, O_{\tau_{m+1}} | (J_{\tau_n}, O_{\tau_n})_{1\le n \le m}, (y_{\tau_n}, \psi_{\tau_n})_{0\le n \le m}, \tilde{y}_{\tau_{m+1}}, \psi_{\tau_{m+1}}\right)\\
     &&\times \frac{ p\left( (J_{\tau_n}, O_{\tau_n})_{1\le n \le m} | (y_{\tau_n}, \psi_{\tau_n})_{0\le n \le m}\right) p\left( \tilde{y}_{\tau_{m+1}}, \psi_{\tau_{m+1}}  | y_{\tau_m}, \psi_{\tau_m}\right) p\left((y_{\tau_n}, \psi_{\tau_n})_{0\le n \le m}\right)}{p\left(J_{\tau_{m+1}}, O_{\tau_{m+1}} | (J_{\tau_n}, O_{\tau_n})_{1\le n \le m} \right)p\left((J_{\tau_n}, O_{\tau_n})_{1\le n \le m}\right)}.\\
\end{eqnarray*}
By using Bayes' rule, we also have
\begin{eqnarray*}
&& p\left((y_{\tau_n}, \psi_{\tau_n})_{0\le n \le m} | (J_{\tau_n}, O_{\tau_n})_{1\le n \le m}\right)\\
&=& \frac{p\left( (J_{\tau_n}, O_{\tau_n})_{1\le n \le m} | (y_{\tau_n}, \psi_{\tau_n})_{0\le n \le m}\right) p\left((y_{\tau_n}, \psi_{\tau_n})_{0\le n \le m}\right)}{p\left((J_{\tau_n}, O_{\tau_n})_{1\le n \le m}\right)}. \\
\end{eqnarray*}
Consequently,
\begin{eqnarray*}
  % \nonumber to remove numbering (before each equation)
    &&p\left((y_{\tau_n}, \psi_{\tau_n})_{0\le n \le m}, \tilde{y}_{\tau_{m+1}}, \psi_{\tau_{m+1}} | (J_{\tau_n}, O_{\tau_n})_{1\le n \le m+1}\right)\\
    &=&p\left((y_{\tau_n}, \psi_{\tau_n})_{0\le n \le m} | (J_{\tau_n}, O_{\tau_n})_{1\le n \le m}\right)\\
     &&\times \frac{ p\left( J_{\tau_{m+1}}, O_{\tau_{m+1}} | (J_{\tau_n}, O_{\tau_n})_{1\le n \le m}, (y_{\tau_n}, \psi_{\tau_n})_{0\le n \le m}, \tilde{y}_{\tau_{m+1}}, \psi_{\tau_{m+1}}\right) p\left( \tilde{y}_{\tau_{m+1}}, \psi_{\tau_{m+1}}  | y_{\tau_m}, \psi_{\tau_m}\right)}{p\left(J_{\tau_{m+1}}, O_{\tau_{m+1}} | (J_{\tau_n}, O_{\tau_n})_{1\le n \le m} \right)}\\
    &=&p\left((y_{\tau_n}, \psi_{\tau_n})_{0\le n \le m} | (J_{\tau_n}, O_{\tau_n})_{1\le n \le m}\right)\\
     &&\times \frac{ p\left(O_{\tau_{m+1}} | (J_{\tau_n})_{1\le n \le m+1}, (O_{\tau_n})_{1\le n \le m}, (y_{\tau_n}, \psi_{\tau_n})_{0\le n \le m}, \tilde{y}_{\tau_{m+1}}, \psi_{\tau_{m+1}}\right) p\left( \tilde{y}_{\tau_{m+1}}, \psi_{\tau_{m+1}}  | y_{\tau_m}, \psi_{\tau_m}\right)}{p\left(O_{\tau_{m+1}} | (J_{\tau_n})_{1\le n \le m+1}, (O_{\tau_n})_{1\le n \le m} \right)}\\
     &&\times \frac{ p\left(J_{\tau_{m+1}} | (J_{\tau_n})_{1\le n \le m}, (O_{\tau_n})_{1\le n \le m}, (y_{\tau_n}, \psi_{\tau_n})_{0\le n \le m}, \tilde{y}_{\tau_{m+1}}, \psi_{\tau_{m+1}}\right)}{p\left(J_{\tau_{m+1}} | (J_{\tau_n})_{1\le n \le m}, (O_{\tau_n})_{1\le n \le m}\right)}.
\end{eqnarray*}
Now, if we assume that the security and the type of trade (summed up in $J_{\tau_{m+1}}$) are independent of the mid-YtB and half bid-ask spread trajectories, then we obtain the following equation:
\begin{eqnarray*}
  % \nonumber to remove numbering (before each equation)
    &&p\left((y_{\tau_n}, \psi_{\tau_n})_{0\le n \le m}, \tilde{y}_{\tau_{m+1}}, \psi_{\tau_{m+1}} | (J_{\tau_n}, O_{\tau_n})_{1\le n \le m+1}\right)\\
     &=&p\left((y_{\tau_n}, \psi_{\tau_n})_{0\le n \le m} | (J_{\tau_n}, O_{\tau_n})_{1\le n \le m}\right)\\
     &&\times \frac{ p\left(O_{\tau_{m+1}} | (J_{\tau_n})_{1\le n \le m+1}, (O_{\tau_n})_{1\le n \le m}, (y_{\tau_n}, \psi_{\tau_n})_{0\le n \le m}, \tilde{y}_{\tau_{m+1}}, \psi_{\tau_{m+1}}\right) p\left( \tilde{y}_{\tau_{m+1}}, \psi_{\tau_{m+1}}  | y_{\tau_m}, \psi_{\tau_m}\right)}{p\left(O_{\tau_{m+1}} | (J_{\tau_n})_{1\le n \le m+1}, (O_{\tau_n})_{1\le n \le m} \right)}.\\
\end{eqnarray*}

Let us notice that in all of the 5 cases considered earlier, we have  $$p\left(O_{\tau_{m+1}} | (J_{\tau_n})_{1\le n \le m+1}, (O_{\tau_n})_{1\le n \le m}, (y_{\tau_n}, \psi_{\tau_n})_{0\le n \le m}, \tilde{y}_{\tau_{m+1}}, \psi_{\tau_{m+1}}\right) = p\left(O_{\tau_{m+1}} | \tilde{y}_{\tau_{m+1}}, \psi_{\tau_{m+1}}, J_{\tau_{m+1}}\right).$$

Therefore,
\begin{eqnarray*}
  % \nonumber to remove numbering (before each equation)
    &&p\left((y_{\tau_n}, \psi_{\tau_n})_{0\le n \le m}, \tilde{y}_{\tau_{m+1}}, \psi_{\tau_{m+1}} | (J_{\tau_n}, O_{\tau_n})_{1\le n \le m+1}\right)\\
     &=&p\left((y_{\tau_n}, \psi_{\tau_n})_{0\le n \le m} | (J_{\tau_n}, O_{\tau_n})_{1\le n \le m}\right)\\
     &&\times \frac{ p\left(O_{\tau_{m+1}} | \tilde{y}_{\tau_{m+1}}, \psi_{\tau_{m+1}}, J_{\tau_{m+1}}\right) p\left( \tilde{y}_{\tau_{m+1}}, \psi_{\tau_{m+1}}  | y_{\tau_m}, \psi_{\tau_m}\right)}{p\left(O_{\tau_{m+1}} | (J_{\tau_n})_{1\le n \le m+1}, (O_{\tau_n})_{1\le n \le m} \right)}.\\
\end{eqnarray*}

Consequently,
\begin{equation}
\label{recu}
p\left((y_{\tau_n}, \psi_{\tau_n})_{0\le n \le m}, \tilde{y}_{\tau_{m+1}}, \psi_{\tau_{m+1}} | (J_{\tau_n}, O_{\tau_n})_{1\le n \le m+1}\right) \propto p\left((y_{\tau_n}, \psi_{\tau_n})_{0\le n \le m} | (J_{\tau_n}, O_{\tau_n})_{1\le n \le m}\right)$$
$$\times p\left(O_{\tau_{m+1}} | \tilde{y}_{\tau_{m+1}}, \psi_{\tau_{m+1}}, J_{\tau_{m+1}}\right) p\left( \tilde{y}_{\tau_{m+1}}, \psi_{\tau_{m+1}}  | y_{\tau_m}, \psi_{\tau_m}\right).
\end{equation}

Although the left-hand side of Eq. \eqref{recu} involves $\tilde{y}_{\tau_{m+1}}$ and not ${y}_{\tau_{m+1}}$, we regard this equation as recursive and the multiplicative term $p\left(O_{\tau_{m+1}} | \tilde{y}_{\tau_{m+1}}, \psi_{\tau_{m+1}}, J_{\tau_{m+1}}\right) p\left( \tilde{y}_{\tau_{m+1}}, \psi_{\tau_{m+1}}  | y_{\tau_m}, \psi_{\tau_m}\right)$ is referred to as the likelihood multiplier.\\

The next step of the reasoning is to compute the above likelihood multiplier in the 5 cases described above: buy and sell D2C trades, buy and sell ``Traded Away'' RFQs, and D2D trades.

\subsubsection{Likelihood recursion: the 5 cases}

The likelihood multiplier $p\left(O_{\tau_{m+1}} | \tilde{y}_{\tau_{m+1}}, \psi_{\tau_{m+1}}, J_{\tau_{m+1}}\right) p\left( \tilde{y}_{\tau_{m+1}}, \psi_{\tau_{m+1}}  | y_{\tau_m}, \psi_{\tau_m}\right)$ in Eq. \eqref{recu} depends on the type of observations. We now detail its computation in the 5 different cases (hereafter, we denote by $\phi$ the probability density function and by $\Phi$ the cumulative distribution function of a $\mathcal{N}(0,1)$ Gaussian variable).

\begin{itemize}
  \item[$J_{\tau_{m+1}} = (i,1)$:] In the case of a D2C trade where a client buys bond $i$ from dealer $D$, we have
  \noindent \begin{eqnarray}
  &&p\left(O_{\tau_{m+1}} | \tilde{y}_{\tau_{m+1}}, \psi_{\tau_{m+1}}, J_{\tau_{m+1}}\right) p\left( \tilde{y}_{\tau_{m+1}}, \psi_{\tau_{m+1}}  | y_{\tau_m}, \psi_{\tau_m}\right)\nonumber\\
  &=& \delta(Y^i_{\tau_{m+1}} = y^i_{\tau_{m+1}} - \psi^i_{\tau_{m+1}} + \epsilon^i_{\tau_{m+1}})\nonumber\\
  &&\times  p\left( (y^{i'}_{\tau_{m+1}})_{1\le i' \neq i \le d}   | (y^{i'}_{\tau_{m}})_{1\le i' \neq i \le d}, y^i_{\tau_{m+1}}+ \epsilon^i_{\tau_{m+1}} = Y^i_{\tau_{m+1}} + \psi^i_{\tau_{m+1}}, \psi^i_{\tau_{m+1}}    \right)\nonumber\\
  && \times p(y^i_{\tau_{m+1}} + \epsilon^i_{\tau_{m+1}} = Y^i_{\tau_{m+1}} + \psi^i_{\tau_{m+1}} | y^i_{\tau_{m}}, \psi^i_{\tau_{m+1}} ) p(\psi_{\tau_{m+1}} | \psi_{\tau_m})\nonumber\\
  &=& \delta(y^i_{\tau_{m+1}}+ \epsilon^i_{\tau_{m+1}} = Y^i_{\tau_{m+1}} + \psi^i_{\tau_{m+1}} )\nonumber\\
  &&\times p\left( (y^{i'}_{\tau_{m+1}})_{1\le i' \neq i \le d}   | (y^{i'}_{\tau_{m}})_{1\le i' \neq i \le d}, y^i_{\tau_{m+1}}+ \epsilon^i_{\tau_{m+1}} = Y^i_{\tau_{m+1}} + \psi^i_{\tau_{m+1}}, \psi^i_{\tau_{m+1}}    \right)\nonumber\\
  &&\times \frac{1}{\sqrt{{\sigma^i}^2 (\tau_{m+1} - \tau_m) + {\sigma_\epsilon^i}^2}} \phi\left(\frac{Y^i_{\tau_{m+1}}+ \psi^i_{\tau_{m+1}}-y^i_{\tau_{m}}}{\sqrt{{\sigma^i}^2 (\tau_{m+1} - \tau_m) + {\sigma_\epsilon^i}^2}}\right) p(\psi_{\tau_{m+1}} | \psi_{\tau_m}). \label{like1}
  \end{eqnarray}

  \item[$J_{\tau_{m+1}} = (i,2)$:] In the case of a D2C trade where a client sells bond $i$ to dealer $D$, we have
  \noindent \begin{eqnarray}
  &&p\left(O_{\tau_{m+1}} | \tilde{y}_{\tau_{m+1}}, \psi_{\tau_{m+1}}, J_{\tau_{m+1}}\right) p\left( \tilde{y}_{\tau_{m+1}}, \psi_{\tau_{m+1}}  | y_{\tau_m}, \psi_{\tau_m}\right)\nonumber\\
  &=& \delta(Y^i_{\tau_{m+1}} = y^i_{\tau_{m+1}} + \psi^i_{\tau_{m+1}}+\epsilon^i_{\tau_{m+1}} )\nonumber\\
  &&\times  p\left( (y^{i'}_{\tau_{m+1}})_{1\le i' \neq i \le d}   | (y^{i'}_{\tau_{m}})_{1\le i' \neq i \le d}, y^i_{\tau_{m+1}}+\epsilon^i_{\tau_{m+1}} = Y^i_{\tau_{m+1}} - \psi^i_{\tau_{m+1}}, \psi^i_{\tau_{m+1}}    \right)\nonumber\\
  && \times p(y^i_{\tau_{m+1}}+\epsilon^i_{\tau_{m+1}} = Y^i_{\tau_{m+1}} - \psi^i_{\tau_{m+1}} | y^i_{\tau_{m}}, \psi^i_{\tau_{m+1}} ) p(\psi_{\tau_{m+1}} | \psi_{\tau_m})\nonumber\\
  &=& \delta(y^i_{\tau_{m+1}}+\epsilon^i_{\tau_{m+1}} =  Y^i_{\tau_{m+1}} - \psi^i_{\tau_{m+1}} )\nonumber\\
  &&\times  p\left( (y^{i'}_{\tau_{m+1}})_{1\le i' \neq i \le d}   | (y^{i'}_{\tau_{m}})_{1\le i' \neq i \le d}, y^i_{\tau_{m+1}}+\epsilon^i_{\tau_{m+1}} = Y^i_{\tau_{m+1}} - \psi^i_{\tau_{m+1}}, \psi^i_{\tau_{m+1}}    \right)\nonumber\\
  &&\times \frac{1}{\sqrt{{\sigma^i}^2 (\tau_{m+1} - \tau_m) + {\sigma_\epsilon^i}^2 }} \phi\left(\frac{Y^i_{\tau_{m+1}}- \psi^i_{\tau_{m+1}}-y^i_{\tau_{m}}}{\sqrt{{\sigma^i}^2 (\tau_{m+1} - \tau_m) + {\sigma_\epsilon^i}^2 }}\right)p(\psi_{\tau_{m+1}} | \psi_{\tau_m}).\label{like2}
  \end{eqnarray}

  \item[$J_{\tau_{m+1}} = (i,3)$:] In the case of a D2C trade where a client buys bond $i$ from another dealer, we have
  \noindent \begin{eqnarray}
  &&p\left(O_{\tau_{m+1}} | \tilde{y}_{\tau_{m+1}}, \psi_{\tau_{m+1}}, J_{\tau_{m+1}}\right) p\left( \tilde{y}_{\tau_{m+1}}, \psi_{\tau_{m+1}}  | y_{\tau_m}, \psi_{\tau_m}\right)\nonumber\\
  &=& \delta(Z^i_{\tau_{m+1}} \le y^i_{\tau_{m+1}} - \psi^i_{\tau_{m+1}} + \epsilon^i_{\tau_{m+1}})\nonumber\\
  &&\times p\left( \tilde{y}_{\tau_{m+1}}, \psi_{\tau_{m+1}}  | y_{\tau_m}, \psi_{\tau_m}\right)\nonumber\\
  &=& \delta(y^i_{\tau_{m+1}} + \epsilon^i_{\tau_{m+1}} \ge Z^i_{\tau_{m+1}} + \psi^i_{\tau_{m+1}} )\nonumber\\
  &&\times p\left( (y^{i'}_{\tau_{m+1}})_{1\le i' \neq i \le d}   | (y^{i'}_{\tau_{m}})_{1\le i' \neq i \le d}, y^i_{\tau_{m+1}} + \epsilon^i_{\tau_{m+1}} \right)\nonumber\\
  &&\times p(y^i_{\tau_{m+1}}+\epsilon^i_{\tau_{m+1}} | y^i_{\tau_{m+1}}+\epsilon^i_{\tau_{m+1}} \ge Z^i_{\tau_{m+1}} + \psi^i_{\tau_{m+1}},  y^i_{\tau_{m}}, \psi^i_{\tau_{m+1}} )\nonumber\\
  &&\times p(y^i_{\tau_{m+1}}+\epsilon^i_{\tau_{m+1}} \ge Z^i_{\tau_{m+1}} + \psi^i_{\tau_{m+1}}| y^i_{\tau_{m}}, \psi^i_{\tau_{m+1}}) p(\psi_{\tau_{m+1}} | \psi_{\tau_m})\nonumber\\
  &=& \delta(y^i_{\tau_{m+1}} + \epsilon^i_{\tau_{m+1}} \ge Z^i_{\tau_{m+1}} + \psi^i_{\tau_{m+1}} )\nonumber\\
  &&\times p\left( (y^{i'}_{\tau_{m+1}})_{1\le i' \neq i \le d}   | (y^{i'}_{\tau_{m}})_{1\le i' \neq i \le d}, y^i_{\tau_{m+1}} + \epsilon^i_{\tau_{m+1}} \right)\nonumber\\
  &&\times p(y^i_{\tau_{m+1}}+\epsilon^i_{\tau_{m+1}} | y^i_{\tau_{m+1}}+\epsilon^i_{\tau_{m+1}} \ge Z^i_{\tau_{m+1}} + \psi^i_{\tau_{m+1}},  y^i_{\tau_{m}}, \psi^i_{\tau_{m+1}} )\nonumber\\
  &&\times \Phi\left(-\frac{Z^i_{\tau_{m+1}} + \psi^i_{\tau_{m+1}} - y^i_{\tau_{m}}}{\sqrt{{\sigma^i}^2 (\tau_{m+1} - \tau_m) + {\sigma_\epsilon^i}^2 }}\right) p(\psi_{\tau_{m+1}} | \psi_{\tau_m}).\label{like3}
  \end{eqnarray}

    \item[$J_{\tau_{m+1}} = (i,4)$:] In the case of a D2C trade where a client sells bond $i$ to another dealer, we have
    \noindent \begin{eqnarray}
  &&p\left(O_{\tau_{m+1}} | \tilde{y}_{\tau_{m+1}}, \psi_{\tau_{m+1}}, J_{\tau_{m+1}}\right) p\left( \tilde{y}_{\tau_{m+1}}, \psi_{\tau_{m+1}}  | y_{\tau_m}, \psi_{\tau_m}\right)\nonumber\\
  &=& \delta(Z^i_{\tau_{m+1}} \ge y^i_{\tau_{m+1}} + \psi^i_{\tau_{m+1}} + \epsilon^i_{\tau_{m+1}} )\nonumber\\
  &&\times  p\left( \tilde{y}_{\tau_{m+1}}, \psi_{\tau_{m+1}}  | y_{\tau_m}, \psi_{\tau_m}\right)\nonumber\\
  &=& \delta(y^i_{\tau_{m+1}} + \epsilon^i_{\tau_{m+1}} \le Z^i_{\tau_{m+1}} - \psi^i_{\tau_{m+1}} )\nonumber\\
  &&\times  p\left( (y^{i'}_{\tau_{m+1}})_{1\le i' \neq i \le d}   | (y^{i'}_{\tau_{m}})_{1\le i' \neq i \le d}, y^i_{\tau_{m+1}} + \epsilon^i_{\tau_{m+1}} \right)\nonumber\\
  &&\times p(y^i_{\tau_{m+1}} + \epsilon^i_{\tau_{m+1}} | y^i_{\tau_{m+1}} + \epsilon^i_{\tau_{m+1}} \le Z^i_{\tau_{m+1}} - \psi^i_{\tau_{m+1}},  y^i_{\tau_{m}}, \psi^i_{\tau_{m+1}} )\nonumber\\
  &&\times p(y^i_{\tau_{m+1}} + \epsilon^i_{\tau_{m+1}} \le Z^i_{\tau_{m+1}} - \psi^i_{\tau_{m+1}} |  y^i_{\tau_{m}}, \psi^i_{\tau_{m+1}}) p(\psi_{\tau_{m+1}} | \psi_{\tau_m})\nonumber\\
  &=& \delta(y^i_{\tau_{m+1}} + \epsilon^i_{\tau_{m+1}} \le Z^i_{\tau_{m+1}} - \psi^i_{\tau_{m+1}} )\nonumber\\
  &&\times  p\left( (y^{i'}_{\tau_{m+1}})_{1\le i' \neq i \le d}   | (y^{i'}_{\tau_{m}})_{1\le i' \neq i \le d}, y^i_{\tau_{m+1}} + \epsilon^i_{\tau_{m+1}} \right)\nonumber\\
  &&\times p(y^i_{\tau_{m+1}} + \epsilon^i_{\tau_{m+1}} | y^i_{\tau_{m+1}} + \epsilon^i_{\tau_{m+1}} \le Z^i_{\tau_{m+1}} - \psi^i_{\tau_{m+1}},  y^i_{\tau_{m}}, \psi^i_{\tau_{m+1}} )\nonumber\\
  &&\times \Phi\left(\frac{Z^i_{\tau_{m+1}} - \psi^i_{\tau_{m+1}} - y^i_{\tau_{m}}}{\sqrt{{\sigma^i}^2 (\tau_{m+1} - \tau_m) + {\sigma_\epsilon^i}^2 }}\right) p(\psi_{\tau_{m+1}} | \psi_{\tau_m}).  \label{like4}
  \end{eqnarray}

  \item[$J_{\tau_{m+1}} = (i,5)$:] In the case of a D2D trade in bond $i$, we have
  \noindent \begin{eqnarray}
  &&p\left(O_{\tau_{m+1}} | \tilde{y}_{\tau_{m+1}}, \psi_{\tau_{m+1}}, J_{\tau_{m+1}}\right) p\left( \tilde{y}_{\tau_{m+1}}, \psi_{\tau_{m+1}}  | y_{\tau_m}, \psi_{\tau_m}\right)\nonumber\\
  &=& \delta(Y^i_{\tau_{m+1}} \in [y^i_{\tau_{m+1}} - \alpha^i_{\tau_{m+1}} + \epsilon^i_{\tau_{m+1}}, y^i_{\tau_{m+1}} + \alpha^i_{\tau_{m+1}} + \epsilon^i_{\tau_{m+1}} ])\nonumber\\
  &&\times p\left( \tilde{y}_{\tau_{m+1}}, \psi_{\tau_{m+1}}  | y_{\tau_m}, \psi_{\tau_m}\right)\nonumber\\
  &=& \delta(Y^i_{\tau_{m+1}} \in [y^i_{\tau_{m+1}} - \alpha^i_{\tau_{m+1}} + \epsilon^i_{\tau_{m+1}}, y^i_{\tau_{m+1}} + \alpha^i_{\tau_{m+1}} + \epsilon^i_{\tau_{m+1}} ])\nonumber\\
  &&\times p\left( (y^{i'}_{\tau_{m+1}})_{1\le i' \neq i \le d}   | (y^{i'}_{\tau_{m}})_{1\le i' \neq i \le d}, y^i_{\tau_{m+1}} + \epsilon^i_{\tau_{m+1}}\right)\nonumber\\
  && \times p(y^i_{\tau_{m+1}} + \epsilon^i_{\tau_{m+1}} \in [Y^i_{\tau_{m+1}} - \alpha^i_{\tau_{m+1}}, Y^i_{\tau_{m+1}} + \alpha^i_{\tau_{m+1}} ] | y^i_{\tau_{m}} ) p(\psi_{\tau_{m+1}} | \psi_{\tau_m})\nonumber\\
  \end{eqnarray}
  \begin{eqnarray}
  &=& \delta(Y^i_{\tau_{m+1}} \in [y^i_{\tau_{m+1}} - \alpha^i_{\tau_{m+1}} + \epsilon^i_{\tau_{m+1}}, y^i_{\tau_{m+1}} + \alpha^i_{\tau_{m+1}} + \epsilon^i_{\tau_{m+1}} ])\nonumber\\
  &&\times p\left( (y^{i'}_{\tau_{m+1}})_{1\le i' \neq i \le d}   | (y^{i'}_{\tau_{m}})_{1\le i' \neq i \le d}, y^i_{\tau_{m+1}} + \epsilon^i_{\tau_{m+1}}\right)\nonumber\\
  && \times \left(\Phi\left(\frac{Y^i_{\tau_{m+1}} + \alpha^i_{\tau_{m+1}} - y^i_{\tau_{m}}}{\sqrt{{\sigma^i}^2 (\tau_{m+1} - \tau_m) + {\sigma_\epsilon^i}^2 }}\right)-\Phi\left(\frac{Y^i_{\tau_{m+1}} - \alpha^i_{\tau_{m+1}} - y^i_{\tau_{m}}}{\sqrt{{\sigma^i}^2 (\tau_{m+1} - \tau_m) + {\sigma_\epsilon^i}^2 }}\right)\right) p(\psi_{\tau_{m+1}} | \psi_{\tau_m}). \label{like5}
  \end{eqnarray}
\end{itemize}

\subsection{Simulation of trajectories by induction}

Let us consider now that we have built a $K$-sample $\left((y^{m}_{n,k},x^{m}_{n,k}, \psi^{m}_{n,k})_{0 \le n \le m}\right)_{1 \le k \le K}$ for a given $m\ge 0$,\footnote{The case $m=0$ corresponds to the initial draw from the prior distribution $\pi$.} and that we want to build $\left((y^{m+1}_{n,k},x^{m+1}_{n,k}, \psi^{m+1}_{n,k})_{0 \le n \le m+1}\right)_{1 \le k \le K}$. The method we use consists of the 6 steps that follow.\\

\paragraph{Step 1: Drawing half bid-ask spreads}

In the first step, we draw $K$ independent points $(\hat{x}_{m+1,k})_{1 \le k \le K}$ with
  $$\forall k \in \{1, \ldots, K\}, \hat{x}_{m+1,k} \sim \mathcal{N}\left( e^{-A(\tau_{m+1}-\tau_m)}\left({x}^{m,1}_{m,k},\ldots, {x}^{m,d}_{m,k}\right)',\Gamma(\tau_{m+1}-\tau_m)\right).$$ Then, we define the associated half bid-ask spreads $(\hat{\psi}_{m+1,k})_{1 \le k \le K}$ by
  $$\forall k \in \{1, \ldots, K \}, \hat{\psi}_{m+1,k} = \left(
                      \begin{array}{c}
                        \Psi^1 \exp(\hat{x}^{1}_{m+1,k}) \\
                        \vdots \\
                        \Psi^d \exp(\hat{x}^{d}_{m+1,k}) \\
                      \end{array}
                    \right).$$

  \paragraph{Step 2: Computing weights}

In the second step, we compute the weights associated with each particle given the observation at time $\tau_{m+1}$. There are 5 cases depending on $J_{\tau_{m+1}}$:
  \begin{itemize}

\item[$J_{\tau_{m+1}} = (i,1)$:] In the case of a D2C trade where a client buys bond $i$ from dealer $D$, we follow Eq. \eqref{like1} and define
  $$\forall k \in \{1,\ldots, K\}, \omega_k = \phi\left(\frac{Y^i_{\tau_{m+1}}+ \hat{\psi}^i_{m+1,k}-y^{m,i}_{m,k}}{\sqrt{{\sigma^i}^2 (\tau_{m+1} - \tau_m) + {\sigma_\epsilon^i}^2}}\right).$$

\item[$J_{\tau_{m+1}} = (i,2)$:] In the case of a D2C trade where a client sells bond $i$ to dealer $D$, we follow Eq. \eqref{like2} and define
  $$\forall k \in \{1,\ldots, K\}, \omega_k = \phi\left(\frac{Y^i_{\tau_{m+1}}- \hat{\psi}^i_{m+1,k}-y^{m,i}_{m,k}}{\sqrt{{\sigma^i}^2 (\tau_{m+1} - \tau_m) + {\sigma_\epsilon^i}^2}}\right).$$

\item[$J_{\tau_{m+1}} = (i,3)$:] In the case of a ``Traded Away'' RFQ where a client buys bond $i$ from another dealer, we follow Eq.~\eqref{like3} and define
  $$\forall k \in \{1,\ldots, K\}, \omega_k = \Phi\left(-\frac{Z^i_{\tau_{m+1}} + \hat{\psi}^i_{m+1,k}-y^{m,i}_{m,k}}{\sqrt{{\sigma^i}^2 (\tau_{m+1} - \tau_m) + {\sigma_\epsilon^i}^2 }}\right).$$

\item[$J_{\tau_{m+1}} = (i,4)$:] In the case of a ``Traded Away'' RFQ where a client sells bond $i$ to another dealer, we follow Eq.~\eqref{like4} and define
  $$\forall k \in \{1,\ldots, K\}, \omega_k = \Phi\left(\frac{Z^i_{\tau_{m+1}} - \hat{\psi}^i_{m+1,k}-y^{m,i}_{m,k}}{\sqrt{{\sigma^i}^2 (\tau_{m+1} - \tau_m) + {\sigma_\epsilon^i}^2 }}\right).$$

\item[$J_{\tau_{m+1}} = (i,5)$:] In the case of a D2D trade, we follow Eq. \eqref{like5} and define
  $$\forall k \in \{1,\ldots, K\}, \omega_k = \Phi\left(\frac{Y^i_{\tau_{m+1}} + \alpha^i_{\tau_{m+1}} -y^{m,i}_{m,k}}{\sqrt{{\sigma^i}^2 (\tau_{m+1} - \tau_m) + {\sigma_\epsilon^i}^2 }}\right) - \Phi\left(\frac{Y^i_{\tau_{m+1}} - \alpha^i_{\tau_{m+1}} -y^{m,i}_{m,k}}{\sqrt{{\sigma^i}^2 (\tau_{m+1} - \tau_m) + {\sigma_\epsilon^i}^2 }}\right).$$

\end{itemize}

  In all these 5 cases, we define weights by
  $$\forall k \in \{1,\ldots, K\}, w_k = \frac{\omega_k}{\sum_{k'=1}^K \omega_{k'}}.$$

  \begin{rem}
  Computing the above weights may be tricky when the arguments of the function $\phi$ or $\Phi$ is very large in absolute value. In that case, classical tricks can be applied such as factoring out the highest/lowest $\omega_k$ and reasoning on ratios, using classical approximations in the tails (for $\Phi$), etc.
  \end{rem}

  \paragraph{Step 3: Resampling} The third step is based on importance sampling with the weights calculated during the second step. In other words, we define a random function $\xi_{m+1} : \{ 1, \ldots, K\} \to \{ 1, \ldots, K\} $ such that:
   \begin{itemize}
     \item $(\xi_{m+1}(k))_k$ are \emph{i.i.d.}
     \item $\forall (j,k) \in \{ 1, \ldots, K\}^2, \mathbb{P}(\xi_{m+1}(k) = j) = w_j$.
   \end{itemize}
   Then, we define $(y^{m+1}_{n,k},x^{m+1}_{n,k}, \psi^{m+1}_{n,k})_{0 \le n \le m, 1 \le k \le K}$ by:
$$\forall n \in \{0, \ldots, m\}, \forall k \in \{1, \ldots, K\}, (y^{m+1}_{n,k},x^{m+1}_{n,k}) = (y^{m}_{n,\xi_{m+1}(k)},x^{m}_{n,\xi_{m+1}(k)}),$$
and
$$\forall n \in \{0, \ldots, m\}, \forall k \in \{1, \ldots, K\}, \psi^{m+1}_{n,k} = \left(
                      \begin{array}{c}
                        \Psi^1 \exp(x^{m+1,1}_{n,k}) \\
                        \vdots \\
                        \Psi^d \exp(x^{m+1,d}_{n,k}) \\
                      \end{array}
                    \right).$$
We also define $(x^{m+1}_{m+1,k})_{1 \le k \le K}$ by
$$\forall k \in \{1, \ldots, K\}, x^{m+1}_{m+1,k} = \hat{x}_{m+1,\xi_{m+1}(k)},$$
and then $(\psi^{m+1}_{m+1,k})_{1 \le k \le K}$ by
$$\forall k \in \{1, \ldots, K\}, \psi^{m+1}_{m+1,k} = \left(
                      \begin{array}{c}
                        \Psi^1 \exp(x^{m+1,1}_{m+1,k}) \\
                        \vdots \\
                        \Psi^d \exp(x^{m+1,d}_{m+1,k}) \\
                      \end{array}
                    \right).$$

At this stage, what remains to be defined is $(y^{m+1}_{m+1,k})_{1 \le k \le K}$. This is the purpose of the following steps.

\paragraph{Step 4: Drawing $\tilde{y}^i_{\tau_{m+1}}$}

The fourth step consists in drawing mid-YtB plus noise for the bond on which we have an observation (a D2C trade, a ``Traded Away'' RFQ, or a D2D trade). As above, there are 5 cases depending on $J_{\tau_{m+1}}$:
  \begin{itemize}

\item[$J_{\tau_{m+1}} = (i,1)$:] In the case of a D2C trade where a client buys bond $i$ from dealer $D$, we follow Eq. \eqref{like1} and introduce
$$\forall k \in \{1,\ldots, K\}, \tilde{y}^i_{m+1,k} = Y^i_{\tau_{m+1}} + \psi^{m+1,i}_{m+1,k}.$$

\item[$J_{\tau_{m+1}} = (i,2)$:] In the case of a D2C trade where a client sells bond $i$ to dealer $D$, we follow Eq. \eqref{like2} and introduce
$$\forall k \in \{1,\ldots, K\}, \tilde{y}^i_{m+1,k} = Y^i_{\tau_{m+1}} - \psi^{m+1,i}_{m+1,k}.$$

\item[$J_{\tau_{m+1}} = (i,3)$:] In the case of a ``Traded Away'' RFQ where a client buys bond $i$ from another dealer, we follow Eq.~\eqref{like3} and draw for each $k \in \{1,\ldots, K\}$ (independently) a value $\tilde{y}^i_{m+1,k}$ from the right-sided truncated Gaussian distribution\footnote{If a random variable $V$ follows a Gaussian distribution $\mathcal{N}(\mu,\sigma^2)$, we say that $V | V > a$ follows a right-sided truncated Gaussian distribution $\mathcal{N}_>(\mu,\sigma^2,a)$.} $\mathcal{N}_>\left(y^{m+1,i}_{m,k}, {\sigma^i}^2 (\tau_{m+1} - \tau_m) + {\sigma_\epsilon^i}^2, Z^i_{\tau_{m+1}} + \psi^{m+1,i}_{m+1,k} \right)$.

\item[$J_{\tau_{m+1}} = (i,4)$:] In the case of a ``Traded Away'' RFQ where a client sells bond $i$ to another dealer, we follow Eq.~\eqref{like4} and draw for each $k \in \{1,\ldots, K\}$ (independently) a value $\tilde{y}^i_{m+1,k}$ from the left-sided truncated Gaussian distribution\footnote{If a random variable $V$ follows a Gaussian distribution $\mathcal{N}(\mu,\sigma^2)$, we say that $V | V < b$ follows a left-sided truncated Gaussian distribution $\mathcal{N}_<(\mu,\sigma^2,b)$.} $\mathcal{N}_<\left(y^{m+1,i}_{m,k}, {\sigma^i}^2 (\tau_{m+1} - \tau_m) + {\sigma_\epsilon^i}^2, Z^i_{\tau_{m+1}} - \psi^{m+1,i}_{m+1,k} \right)$.

\item[$J_{\tau_{m+1}} = (i,5)$:] In the case of a D2D trade, we follow Eq. \eqref{like5} and draw for each $k \in \{1,\ldots, K\}$ (independently) a value $\tilde{y}^i_{m+1,k}$ from the two-sided truncated Gaussian distribution\footnote{If a random variable $V$ follows a Gaussian distribution $\mathcal{N}(\mu,\sigma^2)$, we say that $V | V \in (a,b)$ follows a two-sided truncated Gaussian distribution $\mathcal{N}_{><}(\mu,\sigma^2,a,b)$.} $$\mathcal{N}_{><}\left(y^{m+1,i}_{m,k}, {\sigma^i}^2 (\tau_{m+1} - \tau_m) + {\sigma_\epsilon^i}^2, Y^i_{\tau_{m+1}} - \alpha^i_{\tau_{m+1}}, Y^i_{\tau_{m+1}} + \alpha^i_{\tau_{m+1}} \right).$$

\end{itemize}

\begin{rem}
For drawing values from a left-sided or right-sided truncated Gaussian distribution, one can use for instance Marsaglia's methods (see for instance the Chapter 9 of \cite{delroye}). For drawing values from a two-sided truncated Gaussian distribution, an interesting reference is \cite{chopintrunc}.
\end{rem}

\paragraph{Step 5: Drawing $y^i_{\tau_{m+1}}$}

Once the sample $(\tilde{y}^{i}_{m+1,k})_{1 \le k \le K}$ of mid-YtB plus noise has been computed, we use Eq. \eqref{cond} to get the YtB. More precisely, for each $k \in \{1,\ldots, K\}$, we draw $y^{m+1,i}_{m+1,k}$ from
  $$\mathcal{N}\left(\frac{{\sigma^{i}}^2 (\tau_{m+1} - \tau_m) \tilde{y}^i_{m+1,k}  + {\sigma^{i}_{\epsilon}}^2 y^{m+1,i}_{m,k} }{{\sigma^{i}}^2 (\tau_{m+1} - \tau_m) + {\sigma^{i}_{\epsilon}}^2}, \frac{{\sigma^{i}}^2 (\tau_{m+1} - \tau_m) {\sigma^{i}_{\epsilon}}^2 }{{\sigma^{i}}^2 (\tau_{m+1} - \tau_m) + {\sigma^{i}_{\epsilon}}^2}  \right).$$

  \paragraph{Step 6: Drawing $(y^{i'}_{\tau_{m+1}})_{1\le i' \neq i \le d}$}

  Finally, we need to draw $(y^{m+1,i'}_{m+1,k})_{1 \le k \le K}$ for $i' \neq i$. For that purpose, we draw for all $k \in \lbrace 1, \ldots, K \rbrace$ a vector
  $((y^{m+1,1}_{m+1,k}, \ldots, y^{m+1,i-1}_{m+1,k}, y^{m+1,i+1}_{m+1,k}, \ldots, y^{m+1,d}_{m+1,k})')_{1 \le k \le K} $ from $$\mathcal{N}(\mu_{k|i},\Sigma_{|i} (\tau_{m+1}-\tau_m)),$$
  where $$\mu_{k|i} = \left(
                      \begin{array}{c}
                        y^{m+1,1}_{m,k} \\
                        \vdots \\
                        y^{m+1,i-1}_{m,k} \\
                        y^{m+1,i+1}_{m,k} \\
                        \vdots \\
                        y^{m+1,d}_{m,k} \\
                      \end{array}
                    \right) + (y^{m+1,i}_{m+1,k} - y^{m+1,i}_{m,k}) \left(
                      \begin{array}{c}
                        \rho^{i,1} \frac{\sigma^1}{\sigma^i} \\
                        \vdots \\
                        \rho^{i,i-1} \frac{\sigma^{i-1}}{\sigma^i} \\
                        \rho^{i,i+1} \frac{\sigma^{i+1}}{\sigma^i} \\
                        \vdots \\
                        \rho^{i,d} \frac{\sigma^d}{\sigma^i} \\
                      \end{array}
                    \right),$$
  and $$\Sigma_{|i} = \left(
                      \begin{array}{cccccc}
                         {\sigma^1}^2 & \cdots & \rho^{1,i-1} \sigma^1 \sigma^{i-1} & \rho^{1,i+1} \sigma^1 \sigma^{i+1} & \cdots & \rho^{1,d} \sigma^1 \sigma^{d}  \\
                        \vdots & \ddots & \vdots & \vdots & \ddots & \vdots \\
                        \rho^{i-1,1} \sigma^{i-1} \sigma^1 & \cdots & \rho^{i-1,i-1} \sigma^{i-1} \sigma^{i-1} & \rho^{i-1,i+1} \sigma^{i-1} \sigma^{i+1} & \cdots & \rho^{i-1,d} \sigma^{i-1} \sigma^{d}  \\
                        \rho^{i+1,1} \sigma^{i+1} \sigma^1 & \cdots & \rho^{i+1,i-1} \sigma^{i+1} \sigma^{i-1} & \rho^{i+1,i+1} \sigma^{i+1} \sigma^{i+1} & \cdots & \rho^{i+1,d} \sigma^{i+1} \sigma^{d}  \\
                        \vdots & \ddots & \vdots & \vdots & \ddots & \vdots \\
                         \rho^{d,1} \sigma^d \sigma^1 & \cdots & \rho^{d,i-1} \sigma^d \sigma^{i-1} & \rho^{d,i+1} \sigma^d \sigma^{i+1} & \cdots & {\sigma^d}^2  \\
                      \end{array}
                    \right)$$$$ - \left(
                      \begin{array}{c}
                        \rho^{i,1}\sigma^1 \\
                        \vdots \\
                        \rho^{i,i-1} \sigma^{i-1}\\
                        \rho^{i,i+1} \sigma^{i+1}\\
                        \vdots \\
                        \rho^{i,d} \sigma^d \\
                      \end{array}
                    \right) ( \rho^{i,1}\sigma^1, \ldots, \rho^{i,i-1} \sigma^{i-1}, \rho^{i,i+1} \sigma^{i+1}, \ldots, \rho^{i,d} \sigma^d ).$$

\subsection{Useful outputs}

By following the above steps recursively in $m$, we obtain two interesting outputs:

\begin{itemize}
  \item The first one is, at each time $\tau_m$, an empirical estimation of the distribution of mid-YtBs and half bid-ask spreads through the $K$ particles $(y^{m}_{m,k})_{1\le k \le K}$ and $(\psi^{m}_{m,k})_{1\le k \le K}$.
  \item The second output is related to the \emph{ex-post} estimation of the trajectory of mid-YtBs and half bid-ask spreads. The samples of trajectories $$\left((y^{N}_{n,k}, \psi^N_{n,k})_{0\le n \le N}\right)_{1\le k \le K}$$ provide indeed an empirical estimation of the (posterior) distribution of
$$(y_{\tau_n}, \psi_{\tau_n})_{0\le n \le N} | (J_{\tau_n}, O_{\tau_n})_{1\le n \le N}.$$
However, it is noteworthy that, for this second output, the estimations have been obtained by using a forward-in-time algorithm whereas there is no real arrow of time in the considered estimation problem. In particular, one could find another estimation of the trajectories by considering a prior distribution at time $\tau_N$ and by using a similar SMC algorithm after a time reversal.
\end{itemize}

It is noteworthy that between two observation times, one could continue to diffuse particles by using the dynamics given by Eqs. (\ref{dS}) and (\ref{dx}). In particular, a market maker can easily find an empirical estimation of the distribution of mid-YtBs and half bid-ask spreads at any time he is requested a price by a client.\\

\section{Estimation of the parameters and illustration on corporate bonds}

We now exemplify the use of our algorithm on a small set of European corporate bonds. The idea, here, is just to show how the model could be used and would behave in practice. For confidentiality reason,\footnote{The data has been provided by HSBC France.} we cannot give the estimated value of the parameters but simply illustrate the algorithm on a reduced-size playground.

\subsection{Estimation of the parameters}

In our presentation of the above particle filter algorithm, we considered that the matrices $\Sigma$, $A$, $V$, and the vector $(\Psi^i)_{1 \le i \le d}$ were given. These matrices and this vector drive the YtB and bid-ask spread processes and need to be estimated. However, we are in fact inferring these processes.\\

There are several ways to address this chicken-and-egg problem.\\

The most rigorous one would consist in replacing our sequential Monte-Carlo algorithm by a more complicated algorithm called $\text{SMC}^2$ (see \cite{smc2}). In the SMC$^2$ approach, which is a fully-Bayesian approach, the particles are not only related to $(y_t)_t$ and $(x_t)_t$, but also to the parameters, and we estimate online both a distribution for the mid-YtB, the half bid-ask spreads, and the parameters defining their dynamics (hence the expression ``fully-Bayesian''). In the case of corporate bonds, however, given the large number of assets and the limited liquidity, it is not possible to use an $\text{SMC}^2$ approach.\footnote{This approach was used in \cite{gpindices} for credit indices.}\\

In fact, as far as corporate bonds are concerned, it is reasonable to estimate off-line the matrices $\Sigma$, $A$, $V$, and the vector $(\Psi^i)_{1 \le i \le d}$, for instance before the start of every trading day/week, and to use the resulting estimation over a day/week.\\

For the covariance matrix $\Sigma$, one can assume that the correlation structure and the volatility levels of the YtBs associated with the CBBT mid-prices are the same as those of our mid-YtBs. Therefore, it is reasonable to estimate $\Sigma$ on CBBT mid-price data.\\

As far as the bid-ask spread parameters are concerned, using CBBT bid-ask spreads (\emph{i.e.} quoted spreads) as a proxy for effective bid-ask spreads and carrying out the estimations on the former is not a good idea because CBBT bid and ask prices (or YtBs) are based on streamed (indicative) prices. However, considering that for each trade, the absolute value of the difference between the YtB associated with the trade and the YtB associated with the CBBT mid-price is a proxy of the half bid-ask spread makes sense. With these proxies, one can estimate the matrices $A$, $V$, and the vector $(\Psi^i)_{1 \le i \le d}$. In practice, on our dataset, we found it hard to justify continuity in the bid-ask spread trajectories. As a consequence, we eventually considered a simpler model (see Remark~3) where the half bid-ask spreads are temporally independent and log-normally distributed.\\

\begin{rem}
When it comes to illiquid bonds, it is interesting to build a model that relates our proxy of bid-ask spread to the CBBT bid-ask spread. By doing so, we can have proxies of bid-ask spreads on which to carry out estimations whenever CBBT prices are available.\\
\end{rem}

\subsection{Illustration on a few European corporate bonds}

We took $d=3$ corporate bonds from the same issuer.

The correlation matrix and the volatilities of the YtB of the bonds are estimated historically on the YtB associated with the CBBT prices of the bonds.

We obtained:
\begin{itemize}
\item $\sigma^1 = 0.50\text{bp}\cdot\text{day}^{-\frac12}$,
\item $\sigma^2 = 0.62\text{bp}\cdot\text{day}^{-\frac12}$,
\item $\sigma^3 = 0.69\text{bp}\cdot\text{day}^{-\frac12}$,
\item  $\rho^{1,2} = 0.843$, $\rho^{1,3} = 0.835$ and $\rho^{2,3} = 0.887$.
\end{itemize}

For the diffusion parameters of the variable $x$, we considered, as explained before, a limit case of our framework where the variables $(x_t)_t$ are independent and Gaussian. For the choice of the moments, we matched, for each bond, the mean and the standard deviation of the half bid-ask spreads $\psi_t = e^{x_t}$ to a fraction (here $\frac13$) of the average CBBT half bid-ask spread.

We have:

\begin{itemize}
\item $\mathbb{E}\left[\psi^{1}\right]=\sqrt{\mathbb{V}\left[\psi^{1}\right]} = 0.79 \text{bp}$,
\item $\mathbb{E}\left[\psi^{2}\right]=\sqrt{\mathbb{V}\left[\psi^{2}\right]} = 0.73 \text{bp}$,
\item $\mathbb{E}\left[\psi^{3}\right]=\sqrt{\mathbb{V}\left[\psi^{3}\right]} = 0.65 \text{bp}$.
\end{itemize}

We used our algorithm with $K=10,000$ particles.

In the model, we introduced a noise $\epsilon_t^i$.  For the choice of the standard deviations $(\sigma_\epsilon^i)_{1 \le i \le d}$ we considered, for each bond, a fraction ($5\%$) of its average CBBT bid-ask spread.\\

\begin{figure}[H]
\begin{centering}
\includegraphics[width=0.95\textwidth]{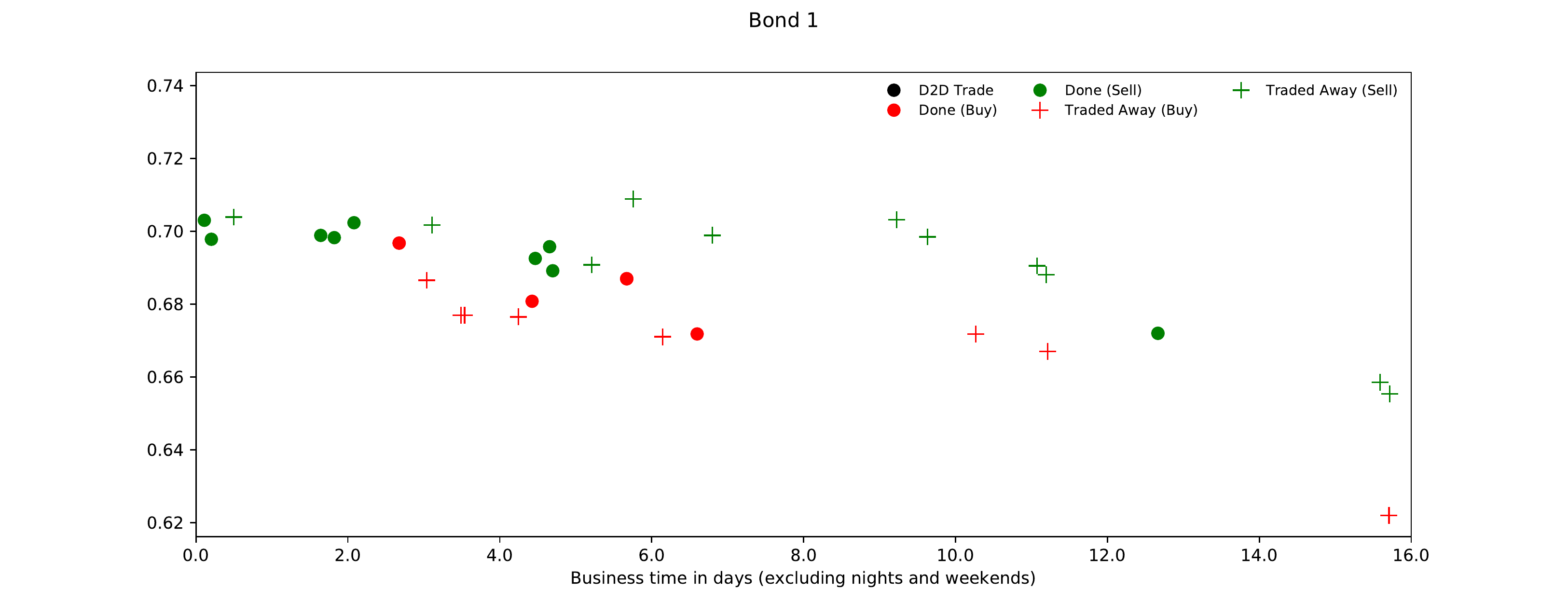}\\
\includegraphics[width=0.95\textwidth]{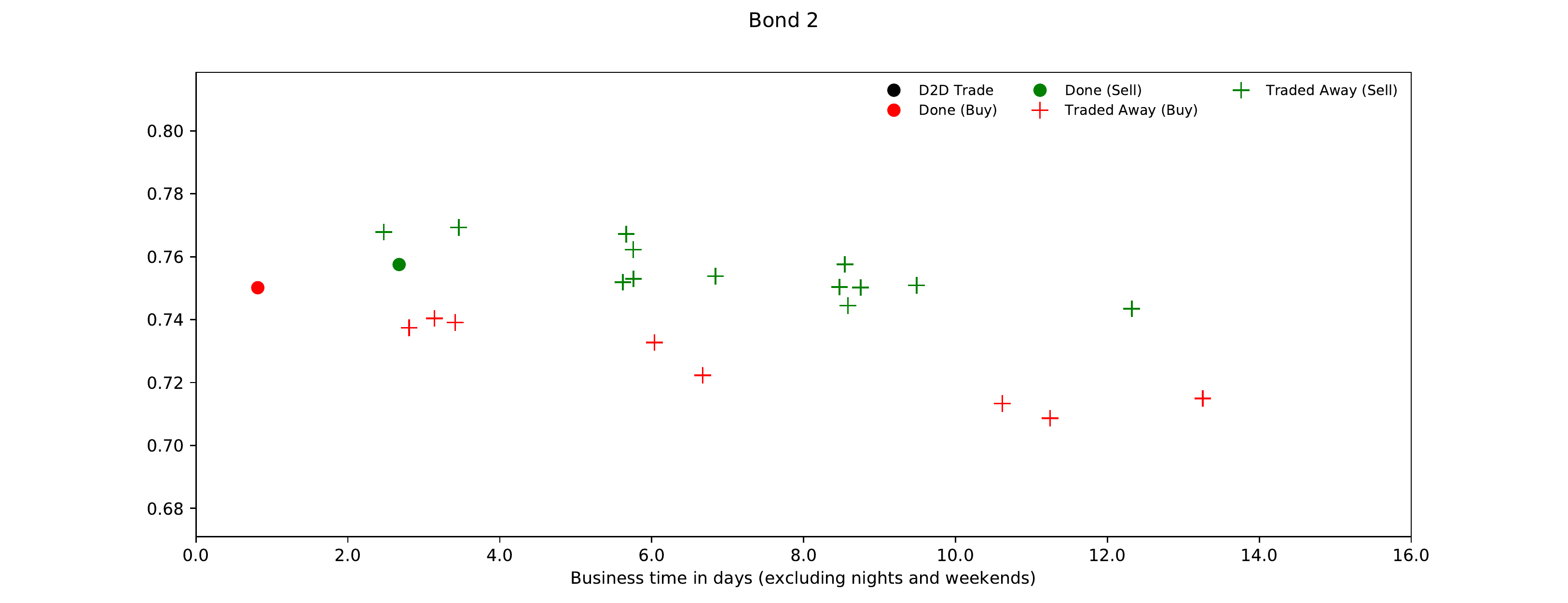}\\
\includegraphics[width=0.95\textwidth]{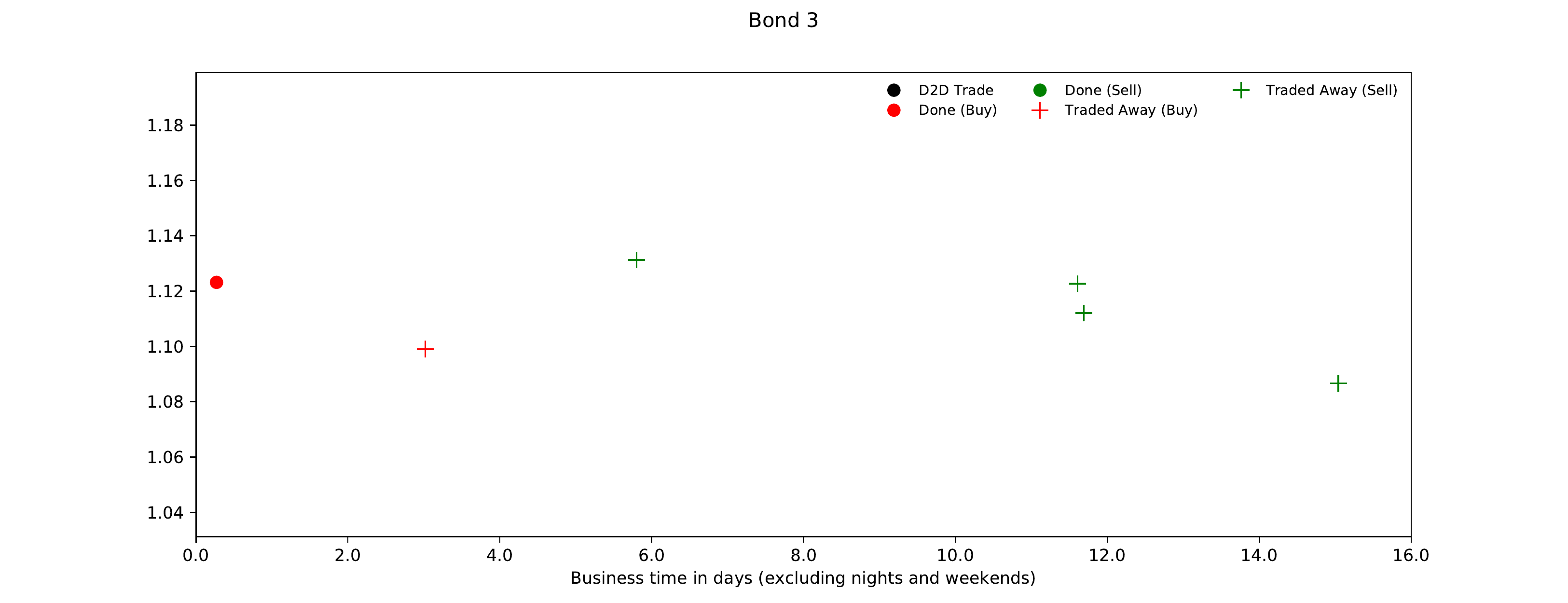}\caption{Observations for the three bonds (unit for the YtB: $\%$)}
\par\end{centering}
\end{figure}

\begin{figure}[H]
\begin{centering}
\includegraphics[width=1.\textwidth]{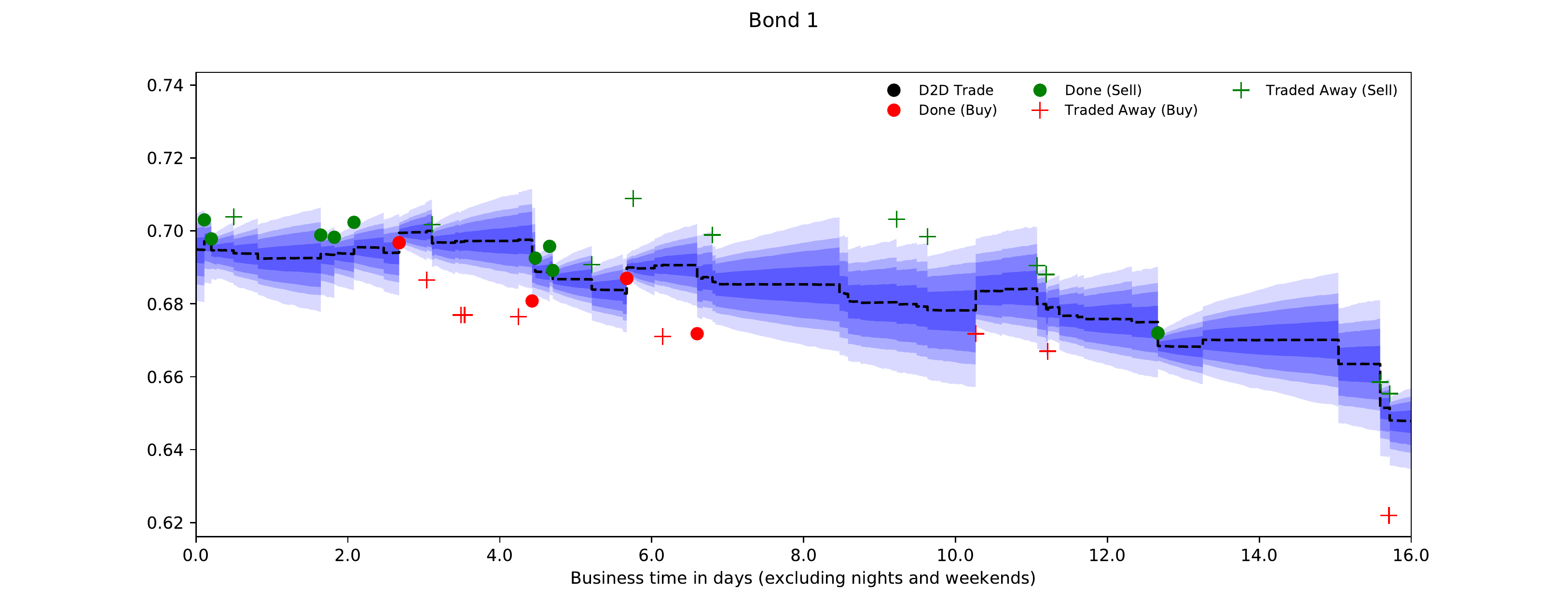}\\
\includegraphics[width=1.\textwidth]{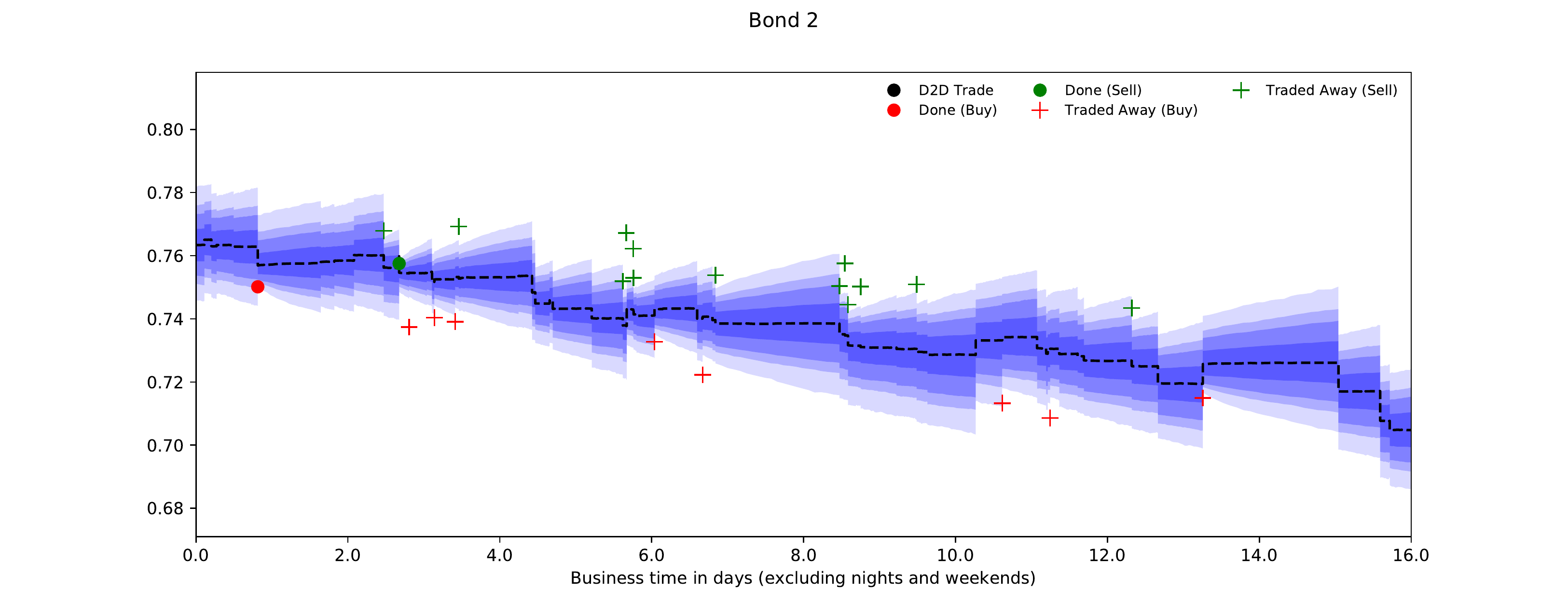}\\
\includegraphics[width=1.\textwidth]{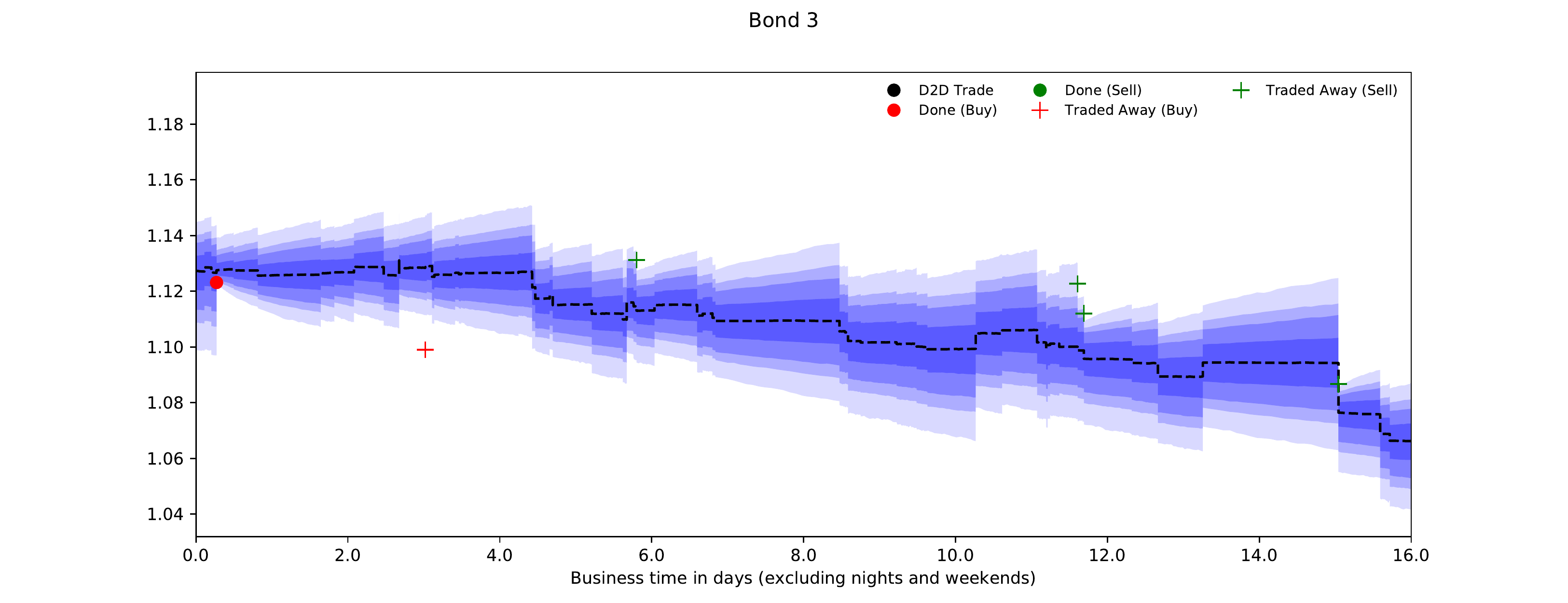}\caption{Estimations (unit for the YtB: $\%$)}
\par\end{centering}
\end{figure}

In order to better understand the quality of the model, we have represented confidence intervals corresponding to quantile envelopes (25\%-75\%, 10\%-90\%, 5\%-95\%, and 1\%-99\%) of the empirical mid-YtB distributions. We see in Figure 2 that, because of the correlation structure, events occurring for one of the assets do enable to better estimate the mid-YtB of other assets. In particular, the algorithm does detect the downward dynamics of Bond 3 in Figure 2 in spite of the small number of observations.

\section{Discussion on the algorithm and possible extensions}

\subsection{Discussion on the algorithm}

\paragraph{From Kalman filter to particle filter}

Kalman filters and their extensions to nonlinear frameworks are the first tools that come to mind for estimating the mid-YtB of bonds given transaction prices. The YtBs associated with buy and sell transactions prices can indeed be regarded as noisy measures of the mid-YtBs, where the distribution of the noises (not necessarily centered) depends on the side of the trades.\\

However, when it comes to including the data coming from RFQs, and in particular the numerous ``Traded Away'' RFQs\footnote{It is interesting to notice that the more numerous the answered RFQs, the more important the dataset. In particular, electronic trading and market making automation help improving the quality of the information.}, Kalman filters are not enough.\\

Particle filters / Sequential Monte-Carlo algorithms constitute an alternative to Kalman filters that are really flexible in terms of underlying models. This is the reason why we chose to develop such an approach.\\

\paragraph{Curse of dimensionality}

Although it is very flexible, the particle filtering approach is often only used to address low-dimensional problems. Particle filters are indeed known to suffer from a form of ``curse of dimensionality'': the number of particles necessary to achieve a given precision when one observes data in dimension $d$ grows exponentially with $d$.\\

In the high-dimensional case of corporate bonds, the ``curse of dimensionality'' that traditionally affects particle filtering / sequential Monte-Carlo approaches must, however, be put into perspective. In illiquid markets, indeed, because transactions seldom occur simultaneously on several securities and because prices diffuse between trades, we are not, by far, in a situation where the observations are really in dimension $d$.\\

In practice, particle filtering approaches on corporate bond data seem to scale far better than expected. We found no problem related to a ``curse of dimensionality'' on a universe of $d=100$ European corporate bonds for instance.\\

\subsection{Extensions}

The model presented in Section 2 and illustrated in Section 3 is an example of model for estimating the mid-YtB (or mid-price) of corporate bonds, based on particle filtering. The modeling framework is in fact very flexible and the model we presented can be modified and extended in many ways.\\

\paragraph{Quoting conventions}

We decided to present a model in which we estimate the trajectory of the mid-YtB. There are in fact several quoting standards and the two most common ones, for investment grade bonds, are the Z-spread and the yield to benchmark, the former being a spread to add to the yield curve associated with the benchmark bond while the latter is a spread to add to the yield to maturity of the benchmark bond. Of course, our model can be used similarly on Z-spreads. As far as high-yield bonds are concerned, they are often quoted in price. In that case, the model has to be adapted, but there is no theoretical difficulty.\\

\paragraph{Distribution of YtB trajectories}

In our model, the mid-YtBs are modelled by Brownian motions. As discussed in Section 2, this is not realistic but it simplifies the exposition. In fact, what is really necessary is to be able to evaluate the weights in Step 2 and to draw the variables involved in Steps 4, 5, and 6. In particular, modeling the mid-YtBs by an Ornstein-Uhlenbeck process slightly complexifies the equations but does not add any technical difficulties.\\

\paragraph{Additional variables}

The goal of the sequential Monte-Carlo algorithm we developed is to estimate (statistically) the mid-YtB and subsequently the mid-price of a set of $d$ corporate bonds. The interest of our multi-dimensional approach is that it can take account of the correlation structure across the $d$ bonds we are interested in, but also between each of the $d$ bonds and other variables that we observe on a more frequent or (nearly-)continuous time basis. It is in fact straightforward to extend the model by adding assets with prices following for instance a Brownian dynamics with a given correlation structure with the other assets.  In the case of European corporate bonds, we can think of adding the spread of iTraxx indices and the price of one or several equity indices (\textit{e.g.} Eurostoxx).\\

In addition to other prices or quotes, it may be interesting to take account of additional variables such as the volume of transactions. In the model we have presented, one considers indeed that the YtB associated with a transaction is independent of its size. Taking account of volumes is not straightforward, but it is possible to imagine a model where the bid-ask spread is a function of the volume and where the particle filter applies to the (dynamic) parameters defining that function.\\

\paragraph{Dealers' inventories}

The algorithm we proposed provides an estimator of the mid-YtBs. As discussed in the introduction, this mid-YtB is endogenously defined in our model by the following property: ``the probability to observe buy or sell trades at any given distance from the reference yield to benchmark only depends on that distance and not on the side''. This property is shared by the reference prices used in most market making models.\footnote{It applies to prices and not to YtBs but the difference can be neglected.} Nevertheless, there are cases where this definition is questionable. In particular, it is difficult to argue that we indeed estimate a mid-YtB when the global position of dealers is very long or when it is very short. In fact, we estimate a mid-YtB skewed by dealers' inventories. If we have an estimate of the inventories or a reason to think that dealers' inventories are skewed, it would be interesting to take this skew into account in order to reduce the bias in the output distribution.\\

\section*{Conclusion}

In this paper, we have proposed a new method based on particle filtering for providing corporate bond market makers with reference prices. The method we propose is fast and very flexible. We believe indeed that it can be extended to include a lot of different features. Furthermore, it can be adapted to numerous other OTC markets.\\

Interestingly, our method provides a complete distribution instead of a single figure for the reference price of bonds. In practice, one could choose to regard the median or the mean of the empirical distribution as the reference price on top of which a market making model is used. However, since the output of Bayesian models is a distribution, it would be interesting to develop a market making quoting model where the reference price is uncertain. In particular, the outcome of such a model would enable to quantify the impact of the uncertainty regarding the value of the security on the optimal bid-ask spreads.


\begin{thebibliography}{10}


\bibitem{avellaneda2008high}
M.~Avellaneda and S.~Stoikov.
\newblock High-frequency trading in a limit order book.
\newblock {\em Quantitative Finance}, 8(3):217--224, 2008.

\bibitem{bis}
Bank for International Settlements.
\newblock Electronic trading in fixed income markets.
\newblock 2016.


\bibitem{cartea2013risk}
{\'A}.~Cartea and S.~Jaimungal.
\newblock Risk metrics and fine tuning of high frequency trading strategies.
\newblock {\em Mathematical Finance}, 25(3):576--611, 2013.


\bibitem{cartea2014buy}
{\'A}.~Cartea, S.~Jaimungal, and J.~Ricci.
\newblock Buy low, sell high: A high frequency trading perspective.
\newblock {\em SIAM Journal on Financial Mathematics}, 5(1):415--444, 2014.


\bibitem{cartea2015algorithmic}
{\'A}.~Cartea, S.~Jaimungal, and J.~Penalva.
\newblock {\em Algorithmic and High-Frequency Trading}.
\newblock Cambridge University Press, 2015.

\bibitem{chopintrunc}
N. Chopin.
\newblock Fast simulation of truncated Gaussian distributions.
\newblock {\em Statistics and Computing}, 21(2), 275-288, 2011.

\bibitem{smc2}
N. Chopin, P. E. Jacob, and O. Papaspiliopoulos.
\newblock SMC2: an efficient algorithm for sequential analysis of state space models.
\newblock {\em Journal of the Royal Statistical Society: Series B (Statistical Methodology)}, 75(3), 397-426, 2013.

\bibitem{delroye}
L. Devroye.
\newblock Non-Uniform Random Variate Generation.
\newblock Springer-Verlag, New York, NY, 1986.

\bibitem{trace}
J. Dick-Nielsen.
\newblock Liquidity biases in TRACE.
\newblock 2009.

\bibitem{fermanian}
J.-D. Fermanian, O. Gu\'eant, and J. Pu.
\newblock The behavior of dealers and clients on the European corporate bond market: the case of Multi-Dealer-to-Client platforms.
\newblock {\em Market Microstructure and Liquidity}, 2(3), 2016.


\bibitem{grossman1988liquidity}
S.~Grossman and M.~Miller.
\newblock Liquidity and market structure.
\newblock {\em The Journal of Finance}, 43(3):617--633, 1988.



\bibitem{gueant2012optimal}
O.~Gu{\'e}ant, C.-A. Lehalle, and J.~Fernandez-Tapia.
\newblock Optimal portfolio liquidation with limit orders.
\newblock {\em SIAM Journal on Financial Mathematics}, 3(1):740--764, 2012.

\bibitem{gueant2013dealing}
O.~Gu{\'e}ant, C.-A. Lehalle, and J.~Fernandez-Tapia.
\newblock Dealing with the inventory risk: a solution to the market making
  problem.
\newblock {\em Mathematics and financial economics}, 7(4):477--507, 2013.

\bibitem{gueant2017optimal}
O.~Gu\'eant
\newblock Optimal market making.
\newblock {\em Applied Mathematical Finance}, 24(2):112--154, 2017

\bibitem{gueantbook}
O.~Gu{\'e}ant.
\newblock The Financial Mathematics of Market Liquidity: from Optimal Execution to Market Making.
\newblock {\em CRC Press, Taylor and Francis}, 2016.

\bibitem{gpindices}
O.~Gu{\'e}ant and J. Pu.
\newblock Mid-price estimation of credit indices: an $\text{SMC}^2$ approach. 2018.

\bibitem{guilbaud2013optimal}
F.~Guilbaud and H.~Pham.
\newblock Optimal high-frequency trading with limit and market orders.
\newblock {\em Quantitative Finance}, 13(1):79--94, 2013.


\bibitem{harris}
L. Harris.
\newblock Transaction costs, trade throughs, and riskless principal trading in
  corporate bond markets.
\newblock 2015.


\bibitem{ho1981optimal}
T.~Ho and H.~Stoll.
\newblock Optimal dealer pricing under transactions and return uncertainty.
\newblock {\em Journal of Financial Economics}, 9(1):47--73, 1981.

\bibitem{ho1983dynamics}
T.~Ho and H.~Stoll.
\newblock The dynamics of dealer markets under competition.
\newblock {\em The Journal of Finance}, 38(4):1053--1074, 1983.

\bibitem{icma}
ICMA.
\newblock The current state and future evolution of the european investment
  grade corporate bond secondary market: perspectives from the market.
\newblock 2014.

\bibitem{iosco}
IOSCO.
\newblock Corporate bond markets: A global perspective.
\newblock 2014.


\bibitem{mck}
{McKinsey \& Company}.
\newblock Corporate bond e-trading: {S}ame game, new playing field.
\newblock 2013.

\bibitem{meucci}
A. Meucci.
\newblock Review of statistical arbitrage, cointegration, and multivariate Ornstein-Uhlenbeck. 2009


\end{thebibliography}
\end{document}